\documentclass[proceedings]{JHEP3}
\usepackage{amsfonts}
\usepackage{amsmath}
\usepackage{epsfig,multicol}

\newbox\mybox

\newcommand\fverb{\setbox\mybox=\hbox\bgroup\verb}
\newcommand\fverbdo{\egroup\medskip\noindent\fbox{\unhbox\mybox}\ }
\newcommand\fverbit{\egroup\item[\fbox{\unhbox\mybox}]}
\conference{PT-symmetric deformations of integrable models}
\abstract{We review recent results on new physical models constructed as ${\mathcal{PT}}$-symmetrical deformations or extensions of different types of integrable models. We present non-Hermitian versions of quantum spin chains, multi-particle systems of Calogero-Moser-Sutherland type and non-linear integrable field equations of Korteweg-de-Vries type. The quantum spin chain discussed is related to the first example in the series of the non-unitary models of minimal conformal field theories. For the Calogero-Moser-Sutherland models we provide three alternative deformations: A complex extension for models related to all types of Coxeter/Weyl groups; models describing the evolution of poles in constrained real valued field equations of non-linear integrable systems and genuine deformations based on antilinearly invariant deformed root systems. Deformations of complex nonlinear integrable field equations of KdV-type are studied with regard to different kinds of ${\mathcal{PT}}$-symmetrical scenarios. A reduction to simple complex quantum mechanical models currently under discussion is presented.}
\title{${\mathcal{PT}}$-symmetric deformations of integrable models}
\author{Andreas Fring \\
Centre for Mathematical Science, City University London, \\
Northampton Square,London EC1V 0HB, UK\\
E-mail: a.fring@city.ac.uk}

\begin{document}

\section{Introduction}

Until fairly\let\thefootnote\relax\footnotetext{%
Invited contribution to the Philosophical Transactions of the Royal Society
A.} recently \cite{Bender:1998ke} non-Hermitian systems have been mostly
viewed as not self-consistent descriptions of dissipative systems. However,
in constrast to the previous misconception it is by now well understood that
Hamiltonians admitting an antilinear symmetry may be used to define
consistent classical, quantum mechanical and quantum field theoretical
systems. Various techniques have been developed to achieve this. Central to
this is construction of metric operators such that certain quantities in the
models can be viewed as physical observables \cite%
{Urubu,Bender:2002vv,Mostafazadeh:2002hb,Bender:2003ve,Mostafazadeh:2004mx,Caliceti:2004xw,CA,Moyal1,ACIso,Mostsyme}%
. In particular, it was found that such type of models models possess real
energy spectra in large sectors in their parameter space, despite being
non-Hermitian. The explanation for this property can be traced back to an
observation made by Wigner more than fifty years ago \cite{EW}, who notices
that operators invariant under antilinear transformations possess either
real eigenvalues or eigenvalues occurring in complex conjugate pairs
depending on whether their eigenfunctions also respect this symmetry or not,
respectively. A very explicit example of such a symmetry is a simultaneous
parity transformation $\mathcal{P}$ and time reversal $\mathcal{T}$. This $%
\mathcal{PT}$-symmetry is trivially verified for instance for Hamiltonian
operators $H$, but less obvious for the corresponding wavefunctions $\psi $
due to the fact that often they are not known explicitly. When 
\begin{equation}
\left[ H,\mathcal{PT}\right] =0\qquad \text{and\qquad }\mathcal{PT}\psi =\psi
\label{PT}
\end{equation}%
hold one speaks of a $\mathcal{PT}$-symmetric system, but when only the
first relation holds one speaks of spontaneously broken $\mathcal{PT}$
-symmetry and when none of the relations in (\ref{PT}) holds of broken $%
\mathcal{PT}$-symmetry. Here we will view the $\mathcal{PT}$-operator in a
wider sense and refer to it loosely as $\mathcal{PT}$ even when it is not
strictly a reflection in time and space, but when it is an antilinear
involution satisfying%
\begin{equation}
\mathcal{PT}\left( \alpha \psi +\beta \phi \right) =\alpha ^{\ast }\mathcal{%
PT}\psi +\beta ^{\ast }\mathcal{PT}\phi \qquad \text{for }\alpha ,\beta \in 
\mathbb{C}\text{, \ \ \ \ \ }\mathcal{PT}^{2}=\mathbb{I}.  \label{anti}
\end{equation}

Very often synonymously used, even though conceptually quite different, are
the notions of quasi-Hermiticity \cite{Dieu,Will,Urubu} and
pseudo-Hermiticity \cite{pseudo1,pseudo2,Mostafazadeh:2002hb}. These
concepts refer more directly to the properties of the metric operator and
their subtle difference is often overlooked, even though this is very
important as they allow for different types of conclusions. In the
quasi-Hermitian case the metric operator is positive and Hermitian, but not
necessarily invertible. It was shown \cite{Dieu,Will,Urubu} that in this
case the existence of a definite metric is guaranteed and the eigenvalues of
the Hamiltonian are real. The pseudo-Hermitian scenario, that is dealing
with an invertible Hermitian, but not necessarily positive metric, is less
appealing as the eigenvalues are only guaranteed to be real but no definite
conclusions can be reached with regard to the existence of a definite
metric. Thus in this latter case the status and consistency of the
corresponding quantum theory remain inconclusive.

Even though some fundamental questions remain partially unanswered, such as
the puzzle concerning the uniqueness of the metric or the question of what
constitutes a good set of ingredients to formulate a consistent physical
theory, the understanding is general in a very mature state. So far it could
be used to revisit some old theories, which had either been discarded as
being non-physical or had considerable gaps in their treatment, and put them
on more solid ground. Another interesting possibility which had opened up
through these studies is the formulation of entirely new models based on
non-Hermitian Hamiltonians which however possess the desired $\mathcal{PT}$%
-symmetry. In other words, one may use the $\mathcal{PT}$-symmetry to deform
or extend previously studied models and thus obtain large sets of entirely
unexplored theories. In principle, this kind of programme can be carried out
in any area of physics. Here we will explore how these ideas can be used in
the context of integrable models. We will not report here on how well
established methods from integrable systems can be applied to study
non-Hermitian quantum mechanical models \cite{DDT}, even though we will
report some scenarios in which they naturally emerge as reduced integrable
systems \cite{CFB}. Instead we present here how these ideas have been used
so far to formulate and study new models previously overlooked as they would
have been regarded as non-physical due to their non-Hermitian nature. We
present results on standard types of integrable models, a quantum
spin-chain, multi-particle systems of Calogero type and nonlinear wave
equations of KdV-type.

The construction principle is fairly simple. Identifying some anti-symmetric
operators $\mathcal{O}$ in the system, we seek a deformation map $\delta
_{\varepsilon }$ of the form

\begin{equation}
\mathcal{PT}:\mathcal{O}\mapsto -\mathcal{O}\quad \Rightarrow \quad \delta
_{\varepsilon }:\mathcal{O}\mapsto -i(i\mathcal{O)}^{\varepsilon },
\label{defPT}
\end{equation}%
with $\varepsilon $ being a deformation parameter such that the non-deformed
model is recovered in the limit $\varepsilon \rightarrow 1$. Alternatively
one can also just add $\mathcal{PT}$-symmetric terms to the original system
and regard them as perturbations.

\section{$\mathcal{PT}$-symmetrically deformed quantum spin chains}

Quantum spin chains constitute a good starting point since, being just
finite matrix models, they can be viewed in many ways as the easiest
integrable models. We present here a model which has been considered first
by von Gehlen \cite{Guenter1}, that is an Ising quantum spin chain in the
presence of a magnetic field in the $z$-direction as well as a longitudinal
imaginary field in the $x$-direction. The corresponding Hamiltonian for a
chain of length $N$ acting on a Hilbert space of the form $(\mathbb{C}%
^{2})^{\otimes N}$ is given by 
\begin{equation}
H(\lambda ,\kappa )=-\frac{1}{2}\sum_{j=1}^{N}(\sigma _{j}^{z}+\lambda
\sigma _{j}^{x}\sigma _{j+1}^{x}+i\kappa \sigma _{j}^{x}),\qquad \lambda
,\kappa \in \mathbb{R}.  \label{H}
\end{equation}%
We used the standard notation for the $2^{N}\times 2^{N}$-matrices $\sigma
_{i}^{x,y,z}=\mathbb{I\otimes I\otimes \ldots \otimes }\sigma
^{x,y,z}\otimes \ldots \otimes \mathbb{I\otimes I}$ with Pauli matrices 
\begin{equation}
\sigma ^{x}=\left( 
\begin{array}{cc}
0 & 1 \\ 
1 & 0%
\end{array}%
\right) ,\qquad \sigma ^{y}=\left( 
\begin{array}{cc}
0 & -i \\ 
i & 0%
\end{array}%
\right) ,\qquad \sigma ^{z}=\left( 
\begin{array}{cc}
1 & 0 \\ 
0 & -1%
\end{array}%
\right) ,  \label{pauli}
\end{equation}%
describing spin 1/2 particles as $i$-th factor acting on the site $i$ of the
chain. This model is of interest as it can be viewed \cite{Cardy:1985yy} as
a perturbation of the $\mathcal{M}_{5,2}$-model in the $\mathcal{M}_{p,q}$%
-series of minimal conformal field theories \cite{BPZ}. It is the simplest
non-unitary model in this infinite class of models, which are all
characterized by the condition $p-q>1$ and whose corresponding Hamiltonians
are all expected to be non-Hermitian. The $\mathcal{PT}$-symmetry of the
model was exploited in \cite{chainOla}.

\subsection{Different versions of $\mathcal{PT}$-symmetry}

Let us first identify the $\mathcal{PT}$-symmetry for the Hamiltonian (\ref%
{H}). Non-Hermitian spin chains have first been studied in this regard in 
\cite{CKW}, where the parity operator $\mathcal{P}^{\prime }:\sigma
_{i}^{x,y,z}\rightarrow \sigma _{N+1-i}^{x,y,z}$ was interpreted quite
literally as a reflection about the center of the chain. Viewing ${\mathcal{T%
}}$ as a standard complex conjugation $\mathcal{P}^{\prime }{\mathcal{T}}$
is then easily identified as a symmetry of the $XXZ$-spin chain Hamiltonian $%
H_{XXZ}$. However, it is seen immediately that this operator is not a
symmetry of the Hamiltonian $H(\lambda ,\kappa )$ in (\ref{H}). Defining
instead \cite{chainOla} the operator%
\begin{equation}
\mathcal{P}:=\prod\limits_{i=1}^{N}\sigma _{i}^{z},\quad \text{with }\quad 
\mathcal{P}^{2}=\mathbb{I}^{\otimes N},  \label{PPP}
\end{equation}%
as an analogue to the parity operator, we may carry out a site-by-site
reflection 
\begin{equation}
\mathcal{P}:(\sigma _{i}^{x},\sigma _{i}^{y},\sigma _{i}^{z})\rightarrow
(-\sigma _{i}^{x},-\sigma _{i}^{y},\sigma _{i}^{z})\quad \text{and\quad }%
\mathcal{T}:(\sigma _{i}^{x},\sigma _{i}^{y},\sigma _{i}^{z})\rightarrow
(\sigma _{i}^{x},-\sigma _{i}^{y},\sigma _{i}^{z}).  \label{eff}
\end{equation}%
It is then easy to verify that this operator is a symmetry of $H(\lambda
,\kappa )$, i.e. we have $\left[ \mathcal{PT},H(\lambda ,\kappa )\right] =0$%
. Clearly this $\mathcal{PT}$-operator acts antilinearly satisfying (\ref%
{anti}) and is therefore a viable candidate for our purposes. In analogy to (%
\ref{PPP}) it is then also suggestive to define 
\begin{equation}
\mathcal{P}_{x}:=\prod\limits_{i=1}^{N}\sigma _{i}^{x}\qquad \text{and\qquad 
}\mathcal{P}_{y}:=\prod\limits_{i=1}^{N}\sigma _{i}^{y},  \label{Pxy}
\end{equation}%
which act as 
\begin{equation}
\mathcal{P}_{x/y}:(\sigma _{i}^{x},\sigma _{i}^{y},\sigma
_{i}^{z})\rightarrow (\pm \sigma _{i}^{x},\mp \sigma _{i}^{y},-\sigma
_{i}^{z}).
\end{equation}%
One can verify that $\left[ \mathcal{P}_{x/y}\mathcal{T},H(\lambda ,\kappa )%
\right] \neq 0$ and $\left[ \mathcal{P}_{x/y}\mathcal{T},H_{XXZ}\right] =0$.
Similar properties can be observed for the non-Hermitian quantum spin chain 
\cite{DeGh} 
\begin{equation}
H_{DG}=\sum_{i=1}^{N}\left( \kappa _{zz}\sigma _{i}^{z}\sigma
_{i+1}^{z}+\kappa _{x}\sigma _{i}^{x}+\kappa _{y}\sigma _{i}^{y}\right) ,
\label{DeGo}
\end{equation}%
with $\kappa _{zz}\in \mathbb{R}$ and $\kappa _{x}$, $\kappa _{y}\in \mathbb{%
C}$. Clearly when $\kappa _{x}$ or $\kappa _{y}\notin \mathbb{R}$ the
Hamiltonian $H_{DG}$ is not Hermitian, but once again one can find suitable
symmetry operators. We notice that $\left[ \mathcal{P}^{\prime }\mathcal{T},H%
\right] \neq 0$, whereas $\left[ \mathcal{PT},H\right] =0$ for $\kappa
_{x},\kappa _{y}\in i\mathbb{R}$ and $\left[ \mathcal{P}_{x/y}\mathcal{T},H%
\right] =0$ for $\kappa _{x/y}\in \mathbb{R},\kappa _{y/x}\in i\mathbb{R}$.

Below we will encounter further ambiguities in the definition of the
antilinear symmetry, which will all manifest in the non-uniqueness of the
metric operator and therefore in the definition of the physics described by
these models. For the Hamiltonians $H_{XXZ}$ and $H_{DG}$ the consequences
of this fact are yet to be explored.

\subsection{The two site model}

\begin{figure}[h]
\centering   \includegraphics[width=6.5cm]{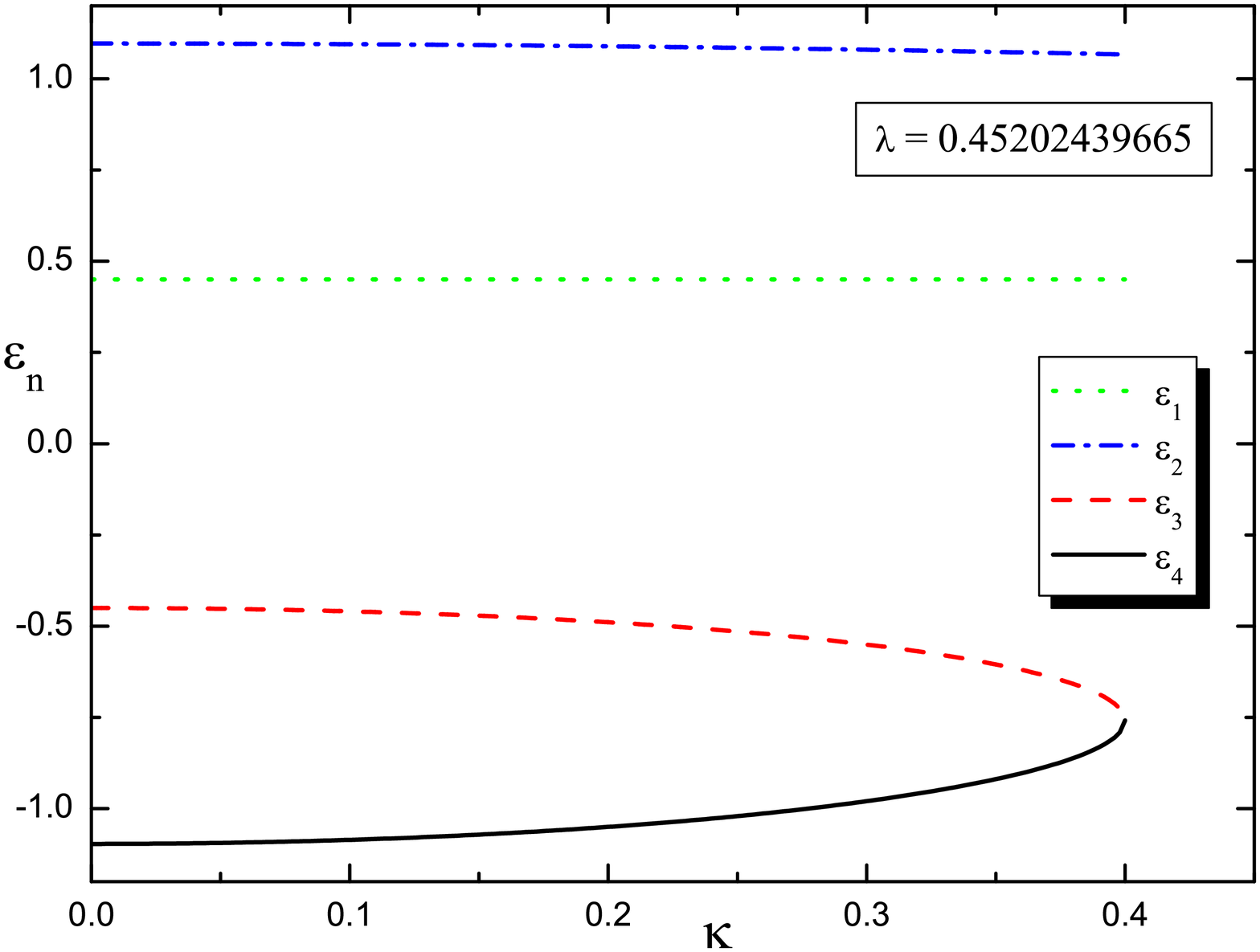} %
\includegraphics[width=6.5cm]{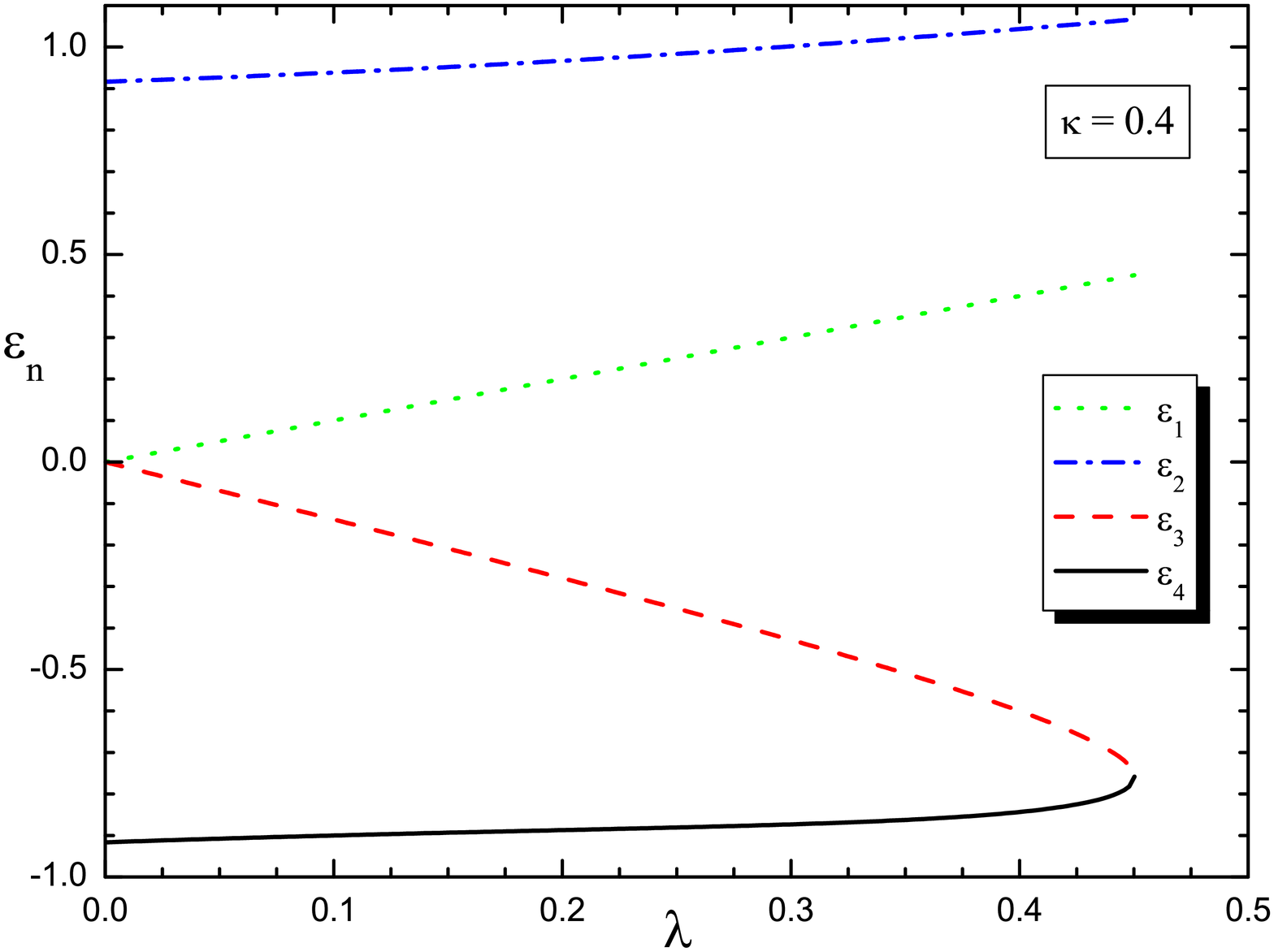}
\caption{Unavoided level crossing: eigenvalues as functions of $\protect%
\lambda $ ($\protect\kappa $) for fixed $\protect\kappa $ ($\protect\lambda $%
) for $H_{2}(\protect\lambda ,\protect\kappa )$.}
\label{F1}
\end{figure}

It is instructive to commence with the simplest example for which all
quantities of interest can be computed explicitly in a very transparent way.
We specify at first the length of the chain to be $N=2$ and without loss of
generality fix the boundary conditions to be periodic $\sigma
_{N+1}^{x}=\sigma _{1}^{x}$. The Hamiltonian (\ref{H}) then acquires the
simple form of a non-Hermitian $(4\times 4)$-matrix%
\begin{eqnarray}
H_{2}(\lambda ,\kappa ) &=&-\frac{1}{2}\left[ \sigma _{z}\otimes \mathbb{I}+%
\mathbb{I}\otimes \sigma _{z}+2\lambda \sigma _{x}\otimes \sigma
_{x}+i\kappa \left( \mathbb{I}\otimes \sigma _{x}+\sigma _{x}\otimes \mathbb{%
I}\right) \right] ,  \label{H2} \\
&=&-\left( 
\begin{array}{rrrr}
-1 & \frac{i\kappa }{2} & \frac{i\kappa }{2} & \lambda \\ 
\frac{i\kappa }{2} & 0 & \lambda & \frac{i\kappa }{2} \\ 
\frac{i\kappa }{2} & \lambda & 0 & \frac{i\kappa }{2} \\ 
\lambda & \frac{i\kappa }{2} & \frac{i\kappa }{2} & -1%
\end{array}%
\right) .  \label{H3}
\end{eqnarray}%
The characteristic polynomial for (\ref{H3}) factorizes into a first and a
third order polynomial such that the eigenvalues acquire a simple analytic
form. Defining the domain 
\begin{equation}
U_{\mathcal{PT}}=\left\{ \lambda ,\kappa :\kappa ^{6}+8\lambda ^{2}\kappa
^{4}-3\kappa ^{4}+16\lambda ^{4}\kappa ^{2}+20\lambda ^{2}\kappa
^{2}+3\kappa ^{2}-\lambda ^{2}-1\leq 0\right\}  \label{upt}
\end{equation}%
in the parameter space, the $\mathcal{PT}$-symmetry is unbroken in the sense
described by (\ref{PT}) when $(\lambda ,$ $\kappa )\in U_{\mathcal{PT}}$.
The four real eigenvalues are evaluated in this case to%
\begin{equation}
\begin{array}{ccc}
\varepsilon _{1}=\lambda , & \varepsilon _{2}=2q^{\frac{1}{2}}\cos \left( 
\frac{\theta }{3}\right) -\frac{\lambda }{3}, & \varepsilon _{3/4}=2q^{\frac{%
1}{2}}\cos \left( \frac{\theta }{3}+\pi \mp \frac{\pi }{3}\right) -\frac{%
\lambda }{3},%
\end{array}%
\end{equation}%
where%
\begin{equation}
\theta =\arccos \left( \frac{r}{q^{3/2}}\right) ,\quad q=\frac{1}{9}\left(
3+4\lambda ^{2}-3\kappa ^{2}\right) ,\quad r=\frac{\lambda }{27}\left(
18\kappa ^{2}+8\lambda ^{2}+9\right) .
\end{equation}%
We depict the eigenvalues in figure \ref{F1} for some fixed $\lambda $ or $%
\kappa $ and varying $\kappa $ or $\lambda $, respectively.

We observe the typical behaviour for $\mathcal{PT}$-symmetric systems,
namely that two eigenvalues start to coincide at the exceptional point \cite%
{CarlExPoint} when $\kappa $ and $\lambda $ are situated on the boundary of $%
U_{\mathcal{PT}}$. Going beyond those values, the $\mathcal{PT}$-symmetry is
spontaneously broken and the two merged eigenvalues develop into a complex
conjugate pair. This is of course a phenomenon prohibited for standard
Hermitian systems by the Wigner--von Neumann non-crossing rule \cite{vNW}.

For the Hamiltonian (\ref{H2}) one can compute explicitly the left $%
\left\vert \Phi _{n}\right\rangle $ and right eigenvectors $\left\vert \Psi
_{n}\right\rangle $, forming a biorthonormal basis 
\begin{equation}
\left\langle \Psi _{n}\right. \left\vert \Phi _{m}\right\rangle =\delta
_{nm},\qquad \text{and\qquad }\sum\nolimits_{n}\left\vert \Phi
_{n}\right\rangle \left\langle \Psi _{n}\right\vert =\mathbb{I}
\end{equation}%
and verify that indeed for the spontaneously broken regime the second
relation in (\ref{PT}) does not hold, see \cite{chainOla} for the concrete
expresssions. We have then all the ingredients to compute the metric
operator $\rho $ and define the inner product $\langle .|.\rangle _{\rho
}:=\langle .|\rho .\rangle $ with regard to which the Hamiltonian (\ref{H2})
is Hermitian%
\begin{equation}
\langle \psi |H\phi \rangle _{\rho }=\langle H\psi |\phi \rangle _{\rho }.
\end{equation}%
Computing the signature $s=(s_{1},s_{2},\ldots ,s_{n})$ from 
\begin{equation}
\mathcal{P}\left\vert \Phi _{n}\right\rangle =s_{n}\left\vert \Psi
_{n}\right\rangle \text{\qquad with }s_{n}=\pm 1  \label{S}
\end{equation}%
we may evaluate the so-called $\mathcal{C}$-operator introduced in \cite%
{Bender:2002vv} 
\begin{equation}
\mathcal{C}:=\sum\nolimits_{n}s_{n}\left\vert \Phi _{n}\right\rangle
\left\langle \Psi _{n}\right\vert ,  \label{C}
\end{equation}%
and hence the metric operator $\rho $, which also relates the Hamiltonian to
its conjugate 
\begin{equation}
\rho :=\mathcal{PC},\mathcal{\qquad }H^{\dagger }\rho =\rho H.  \label{rel}
\end{equation}%
The explicit expressions can be found in\ \cite{chainOla}, from which one
can verify explicitly that the metric operator is Hermitian, positive and
invertible. Thus the Hamiltonian (\ref{H2}) is quasi-Hermitian as well as
pseudo-Hermitian. From the expression for $\rho $ we can also obtain the
so-called Dyson map $\eta $ \cite{Dyson}, by taking the positive square root 
$\eta =\sqrt{\rho }$. This operator serves to construct an isospectral
Hermitian counterpart $h$ to $H$ by its adjoint action. For (\ref{H2}) we
find 
\begin{equation}
h_{2}(\lambda ,\kappa )=\eta H_{2}(\lambda ,\kappa )\eta
^{-1}=\sum\limits_{s=x,y,z}\nu _{s}\sigma _{s}\otimes \sigma _{s}+\mu
_{z}(\sigma _{z}\otimes \mathbb{I}+\mathbb{I}\otimes \sigma _{z}).
\label{eta}
\end{equation}%
The constants $\nu _{x}$, $\nu _{y}$, $\nu _{z}$ and $\mu _{z}$ can be found
in \cite{chainOla}.

\subsection{Perturbative computation for the N-site model}

It is clear that when proceeding to longer spin chains it becomes
increasingly complex to compute the above mentioned operators such that
exact computation become less transparent and can only be carried out with
great effort. However, we may also gain considerable insight by resorting to
a perturbative analysis. For this purpose we separate the Hamiltonian into
its Hermitian and non-Hermitian part as $H(\lambda ,\kappa )=h_{0}(\lambda
)+i\kappa h_{1}$, where $h_{0}$ and $h_{1}$ are both Hermitian, with $\kappa 
$ being a real coupling constant as for instance introduced in (\ref{H}).
Assuming next that the inverse of the metric exists and that it can be
parameterized as $\rho =e^{q}$, the second equation in (\ref{rel}) can be
written as 
\begin{equation}
H^{\dagger }=e^{q}He^{-q}=H+[q,H]+\frac{1}{2}[q,[q,H]]+\frac{1}{3!}%
[q,[q,[q,H]]]+\cdots  \label{per}
\end{equation}%
Presuming further that the metric can be perturbatively expanded as 
\begin{equation}
q=\sum_{k=1}^{\infty }\kappa ^{2k-1}q_{2k-1},  \label{eta12}
\end{equation}%
we obtain the following equations order by order in $\kappa $ 
\begin{eqnarray}
\lbrack h_{0},q_{1}] &=&2ih_{1},  \label{c1} \\
\lbrack h_{0},q_{3}] &=&\frac{i}{6}[q_{1},[q_{1},h_{1}]],  \label{c3} \\
\lbrack h_{0},q_{5}] &=&\frac{i}{6}[q_{1},[q_{3},h_{1}]]+\frac{i}{6}%
[q_{3},[q_{1},h_{1}]]-\frac{i}{360}[q_{1},[q_{1},[q_{1},[q_{1},h_{1}]]]].~~
\label{c5}
\end{eqnarray}%
It is clear from (\ref{c1})-(\ref{c5}) that at each order one new unknown
quantity enters the computation for which we can solve our equations, i.e.
in (\ref{c1}) we solve for $q_{1}$ for known $h_{0}$ and $h_{1}$, in (\ref%
{c3}) for $q_{3}$, in (\ref{c5}) for $q_{5}$, etc. This process can be
continued up to any desired order of precision, see \cite{CA} for further
general details on perturbation theory.

Proceeding in this manner we compute the Dyson operator as described above
and determine the Hermitian counterpart. For $N=3$ we obtain%
\begin{equation}
h=\mu _{xx}^{3}S_{xx}^{3}+\mu _{yy}^{3}S_{yy}^{3}+\mu
_{zz}^{3}S_{zz}^{3}+\mu _{z}^{3}S_{z}^{3}+\mu _{xxz}^{3}S_{xxz}^{3}+\mu
_{yyz}^{3}S_{yyz}^{3}+\mu _{zzz}^{3}S_{zzz}^{3},
\end{equation}%
where for convenience we introduced a new notation%
\begin{equation}
S_{a_{1}a_{2}\ldots a_{p}}^{N}:=\sum_{k=1}^{N}\sigma _{k}^{a_{1}}\sigma
_{k+1}^{a_{2}}\ldots \sigma _{k+p-1}^{a_{p}},~~\text{for}%
~~a_{i}=x,y,z,u;~i=1,\ldots ,p\leq N.  \label{matrices}
\end{equation}%
The coefficients $\mu _{xx},\ldots ,\mu _{zzz}$ are real functions of the
couplings $\lambda $ and $\kappa $. We denote here $\sigma ^{u}=\mathbb{I}$
to allow for the possibility of non-local, i.e. not nearest neighbour,
interactions. In fact, they do occur when we increase the length of the
chain by one site. For $N=4$ we compute 
\begin{eqnarray}
h &=&\mu _{xx}^{4}S_{xx}^{4}+\nu _{xx}^{4}\mathbf{S}_{xux}^{4}+\mu
_{yy}^{4}S_{yy}^{4}+\nu _{yy}^{4}\mathbf{S}_{yuy}^{4}+\mu
_{zz}^{4}S_{zz}^{4}+\nu _{zz}^{4}\mathbf{S}_{zuz}^{4}+\mu
_{z}^{4}S_{z}^{4}~~~~~  \notag \\
&&+\mu _{xxz}^{4}(S_{xxz}^{4}+S_{zxx}^{4})+\mu _{xzx}^{4}S_{xzx}^{4}+\mu
_{yyz}^{4}(S_{yyz}^{4}+S_{zyy}^{4})+\mu _{yzy}^{4}S_{yzy}^{4}  \notag \\
&&+\mu _{zzz}^{4}S_{zzz}^{4}+\mu _{xxxx}^{4}S_{xxxx}^{4}+\mu
_{yyyy}^{4}S_{yyyy}^{4}+\mu _{zzzz}^{4}S_{zzzz}^{4}+\mu
_{xxyy}^{4}S_{xxyy}^{4}  \notag \\
&&+\mu _{xyxy}^{4}S_{xyxy}^{4}+\mu _{zzyy}^{4}S_{zzyy}^{4}+\mu
_{zyzy}^{4}S_{zyzy}^{4}+\mu _{xxzz}^{4}S_{xxzz}^{4}+\mu
_{xzxz}^{4}S_{xzxz}^{4}.\;\;  \label{h4lk}
\end{eqnarray}%
We observe that the first non-local interaction terms proportional to $%
S_{xux}^{4}$, $S_{yuy}^{4}$ and $S_{zuz}^{4}$ emerge in this model. Thus we
encounter a very typical feature of non-Hermitian $\mathcal{PT}$-symmetric
Hamiltonian systems, whereas the non-Hermitian Hamiltonian is fairly simple
its Hermitian isospectral counterpart is quite complicated involving
non-nearest neighbour interactions. An additional feature not present for
chains of smaller length is the fact that some of the $\lambda $-dependence
of the coefficients $\mu _{xx},\ldots ,\mu _{xzxz}$ is no longer polynomial
and gives rise to singularities.

This models exhibits the basic feature, but clearly there is plenty of scope
left for further analysis. More explicit analytic formulae should be
computed for $\eta $, $\rho $ and $h$ for longer chains, models with higher
spin values should be considered and further members of the class belonging
to the perturbed $\mathcal{M}_{p,q}$-series of minimal conformal field
theories should be studied. Interesting recent results on other
non-Hermitian quantum spin chains may be found in \cite{Andrei,Georgi}.

\section{$\mathcal{PT}$-symmetrically deformed Calogero type models}

$\mathcal{PT}$-deformed versions of multi-particle systems of Calogero type
have been obtained so far in three quite different ways, as simple
extensions, as constrained field equations or as genuine deformations.

\subsection{Extended Calogero-Moser-Sutherland models}

The most direct and simplest way to obtain $\mathcal{PT}$-symmetrically
extended versions of Calogero-Moser Sutherland models is to add a $\mathcal{%
PT}$-symmetric term to the original model as proposed in \cite{AF} 
\begin{equation}
\mathcal{H}_{\mathcal{PT}CMS}=\frac{1}{2}p^{2}+\frac{1}{2}%
\sum\limits_{\alpha \in \Delta }g_{\alpha }^{2}V(\alpha \cdot q)+\frac{i}{2}%
\tilde{g}_{\alpha }f(\alpha \cdot q)\alpha \cdot p,  \label{af}
\end{equation}%
with coupling constants $g,\tilde{g}\in \mathbb{R}$, canonical variables $%
q,p\in \mathbb{R}^{\ell +1}$ for an $(\ell +1)$-dimensional representation
of the roots $\alpha $ of some arbitrary root system $\Delta $, which is
left invariant under the Coxeter group. The potential may take on different
forms $V(x)=f^{2}(x)$ defined by means of the function $f(x)=1/x$, $%
f(x)=1/\sinh x$ or $f(x)=1/\sin x$. The model in (\ref{af}) is a
generalization of an extension of the $A_{\ell }$ and $B_{\ell }$-Calogero
model, i.e. $f(x)=1/x$, for a specific representation of the roots as
suggested in \cite%
{Basu-Mallick:2001ce,Basu-Mallick:2003pt,Basu-Mallick:2004ye}. The $\mathcal{%
PT}$-symmetry of $\mathcal{H}_{\mathcal{PT}CMS}$ is easily verified. In \cite%
{AF} it was shown that for $f(x)=1/x$ one may re-write the Hamiltonian in (%
\ref{af}), such that it becomes the standard Hermitian Calogero Hamiltonian
with shifted momenta $p\rightarrow p+i\mu $ 
\begin{equation}
\mathcal{H}_{\mathcal{PT}CMS}=\frac{1}{2}(p+i\mu )^{2}+\frac{1}{2}%
\sum\limits_{\alpha \in \Delta }\hat{g}_{\alpha }^{2}V(\alpha \cdot q),
\label{shift}
\end{equation}%
where $\mu :=1/2\sum\nolimits_{\alpha \in \Delta }\tilde{g}_{\alpha
}f(\alpha \cdot q)\alpha $ and the coupling constants have been redefined to 
$\hat{g}_{\alpha }^{2}:=g_{s}^{2}+\alpha _{s}^{2}\tilde{g}_{s}^{2}$ for $%
\alpha \in \Delta _{s}$ and $\hat{g}_{\alpha }^{2}:=g_{l}^{2}+\alpha _{l}^{2}%
\tilde{g}_{l}^{2}$ for $\alpha \in \Delta _{l}$, where $\Delta _{l}$ and $%
\Delta _{s}$ refer to the root system of the long and short roots,
respectively. This manipulation is based on the not obvious identity $\mu
^{2}=\alpha _{s}^{2}\tilde{g}_{s}^{2}\sum\nolimits_{\alpha \in \Delta
_{s}}V(\alpha \cdot q)+\alpha _{l}^{2}\tilde{g}_{l}^{2}\sum\nolimits_{\alpha
\in \Delta _{l}}V(\alpha \cdot q)$, which is only valid for rational
potentials. Even then it has not been proven yet in a case independent
manner, but verified for many examples on a case-by-case basis \cite{AF}.

For the rational potential it is straightforward to obtain the Dyson map $%
\eta =e^{-q\cdot \mu }$, which relates the standard Hermitian Calogero model
to the non-Hermitian model (\ref{af}) by an adjoint action $\mathcal{H}_{%
\mathcal{PT}C}=\eta ^{-1}\mathcal{H}_{C}\eta $. The integrability of the
rational version of $\mathcal{H}_{\mathcal{PT}CMS}$ follows then from the
existence of the Lax pair $L_{C}$ and $M_{C}$ obeying the Lax equation $\dot{%
L}_{\mathcal{PT}C}=\left[ L_{\mathcal{PT}C},M_{\mathcal{PT}C}\right] $,
which maybe obtained from the standard Calogero Lax pair \cite{OP2} as $L_{%
\mathcal{PT}C}(p)=\eta ^{-1}L_{C}(p)\eta =$ $L_{C}(p+i\mu )$ and $M_{%
\mathcal{PT}C}(p)=\eta ^{-1}M_{C}(p)\eta =$ $M_{C}(p+i\mu )$. Expanding the
shifted kinetic term in (\ref{shift}) we obtain 
\begin{equation}
\mathcal{H}_{\mathcal{PT}CMS}=\frac{1}{2}p^{2}+\frac{1}{2}%
\sum\limits_{\alpha \in \Delta }\hat{g}_{\alpha }^{2}V(\alpha \cdot q)+i\mu
\cdot p-\frac{1}{2}\mu ^{2}.  \label{int}
\end{equation}%
By the reasoning provided, it follows that this model is integrable for all
of the above stated potential, whereas the model without the $\mu ^{2}$-term
is only integrable for rational potentials.

\subsection{From constraint field equations to $\mathcal{PT}$-deformed
Calogero models}

Another more surprising way to obtain particle systems of complex Calogero
type arises from considering real valued field solutions for some nonlinear
equations. Making an Ansatz in form of a real valued field 
\begin{equation}
u(x,t)=\frac{\lambda }{2}\sum\limits_{k=1}^{\ell }\left( \frac{i}{x-z_{k}(t)}%
-\frac{i}{x-z_{k}^{\ast }(t)}\right) ,\qquad \lambda \in \mathbb{R},
\label{re}
\end{equation}%
it was shown more than thirty years ago \cite{Chen,JoshLondon} that this
constitutes an $\ell $-soliton solution for the Benjamin-Ono equation 
\begin{equation}
u_{t}+uu_{x}+\lambda Hu_{xx}=0,  \label{BO}
\end{equation}%
with $Hu(x)$ denoting the Hilbert transform $Hu(x)=\frac{P}{\pi }%
\int_{-\infty }^{\infty }\frac{u(x)}{z-x}dz$, provided the poles $z_{k}$ in (%
\ref{re}) obey the \emph{complex} $A_{\ell }$-Calogero equation of motion 
\begin{equation}
\ddot{z}_{k}=\frac{\lambda ^{2}}{2}\sum\limits_{j\neq
k}(z_{j}-z_{k})^{-3},\qquad z_{k}\in \mathbb{C}.  \label{const}
\end{equation}%
Clearly for different types of nonlinear equations the constraining equation
might be of a more complicated form. However, we may consistently impose
additional constraints by making use of the following theorem of Airault,
McKean and Moser \cite{AMM}:

\emph{Given a Hamiltonian $H(x_{1},\ldots ,x_{n},\dot{x}_{1},\ldots ,\dot{x}%
_{n})$ with flow 
\begin{equation}
x_{i}={\partial H}/{\partial \dot{x}_{i}}\qquad \text{and\qquad }\ddot{x}%
_{i}=-{\partial H}/{\partial x_{i}}\qquad i=1,\ldots ,n
\end{equation}%
and conserved charges $I_{j}$ in involution with $H$, i.e. vanishing Poisson
brackets $\{I_{j},H\}=0$. Then the locus of \texttt{grad} $I=0$ is invariant
with regard to time evolution. Thus it is permitted to restrict the flow to
that locus provided it is not empty.}

Making now the Ansatz 
\begin{equation}
v(x,t)=\lambda \sum\limits_{k=1}^{\ell }{(x-z_{k}(t))^{-2}},\qquad \lambda
\in \mathbb{R}
\end{equation}%
one can show that this solves the Boussinesq equation 
\begin{equation}
v_{tt}=a(v^{2})_{xx}+bv_{xxxx}+v_{xx}\qquad a,b\in \mathbb{R}.  \label{bou}
\end{equation}%
if and only if $b=1/12$, $\lambda =-a/2$ and the poles $z_{k}$ obeys the
constraining equations 
\begin{eqnarray}
\ddot{z}_{k} &=&2\sum\limits_{j\neq k}(z_{j}-z_{k})^{-3}\qquad \quad
\Leftrightarrow \quad \ddot{z}_{k}=-\frac{\partial H_{Cal}}{\partial z_{i}},
\label{equom} \\
\dot{z}_{k}^{2} &=&1-\sum\limits_{j\neq k}(z_{j}-z_{k})^{-2}\qquad
\Leftrightarrow \quad \mathtt{grad}(I_{3}-I_{1})=0.  \label{con}
\end{eqnarray}%
Here $I_{3}=\sum_{j=1}^{\ell }[\dot{z}_{j}^{3}/3+\sum\nolimits_{k\neq j}\dot{%
z}_{j}(z_{j}-z_{k})^{2}]$ and $I_{1}=\sum_{j=1}^{\ell }\dot{z}_{j}$ are two
conserved charges in the $A_{\ell }$-Calogero model. Thus in comparison with
the previous example (\ref{re})-(\ref{BO}) we have to satisfy an additional
constraints (\ref{con}) besides the equations of motion of the $A_{\ell }$%
-Calogero model. However, according to the above theorem this is still a
consistent system of equations, provided the equations (\ref{equom}) and (%
\ref{con}) possess any non-trivial solution. Only very few solutions have
been found so far. The simplest two-particle solution was already reported
in \cite{AMM} 
\begin{equation}
z_{1}=\kappa +\sqrt{(t+\tilde{\kappa})^{2}+1/4},\qquad z_{2}=\kappa -\sqrt{%
(t+\tilde{\kappa})^{2}+1/4}.
\end{equation}%
In this case the Boussinesq solution acquires the form 
\begin{equation}
v(x,t)=2\lambda \frac{(x-\kappa )^{2}+(t+\tilde{\kappa})^{2}+1/4}{[(x-\kappa
)^{2}-(t+\tilde{\kappa})^{2}-1/4]^{2}}.  \label{v}
\end{equation}%
Note that $v(x,t)$ is still a real solution. However, without any
complication we may change $\kappa $ and $\tilde{\kappa}$ to be purely
imaginary in which case, and only in this case, (\ref{v}) becomes a solution
for the $\mathcal{PT}$-symmetric equation (\ref{bou}) in the sense that $%
\mathcal{PT}:x\rightarrow -x,t\rightarrow -t$ and $v\rightarrow v$. A three
particle solution was reported in \cite{Assis:2009gt}, which exhibits an
interesting solitonic behaviour in the complex plane. In that case no real
solution could be found and once again one was forced to consider complex
particle systems. For more particles, different types of algebras or other
types of nonlinear equations these investigations have not been carried out
yet.

\subsection{Deformed Calogero-Moser-Sutherland models}

Let us now consider the $CMS$-models with an additional confining potential 
\begin{equation}
\mathcal{H}_{\mathcal{PT}CMS}^{\prime }=\frac{p^{2}}{2}+\frac{m^{2}}{16}%
\sum_{\alpha \in \Delta _{s}}(\alpha \cdot \tilde{q})^{2}+\frac{1}{2}%
\sum\limits_{\alpha \in \Delta }g_{\alpha }V(\alpha \cdot \tilde{q}%
),\,\,~~~m,g_{\alpha }\in \mathbb{R},  \label{defCMS}
\end{equation}%
and also deforme the coordinates $q\rightarrow \tilde{q}$. Considering at
first the $A_{2}$-case for a standard three dimensional representation for
the simple $A_{2}$-roots $\alpha _{1}=\{1,-1,0\}$, $\alpha _{2}=\{0,1,-1\}$,
we deformed the coordinates as 
\begin{eqnarray}
q_{1} &\rightarrow &\tilde{q}_{1}=q_{1}\cosh \varepsilon ~+i\sqrt{3}%
(q_{2}-q_{3})\sinh \varepsilon \\
q_{2} &\rightarrow &\tilde{q}_{2}=q_{2}\cosh \varepsilon ~+i\sqrt{3}%
(q_{3}-q_{1})\sinh \varepsilon \\
q_{3} &\rightarrow &\tilde{q}_{3}=q_{3}\cosh \varepsilon ~+i\sqrt{3}%
(q_{1}-q_{2})\sinh \varepsilon
\end{eqnarray}%
such that the relevant terms in the potential become 
\begin{eqnarray}
\alpha _{1}\cdot \tilde{q} &=&q_{12}\cosh \varepsilon -\frac{i}{\sqrt{3}}%
(q_{13}+q_{23})\sinh \varepsilon , \\
\alpha _{2}\cdot \tilde{q} &=&q_{23}\cosh \varepsilon -\frac{i}{\sqrt{3}}%
(q_{21}+q_{31})\sinh \varepsilon , \\
(\alpha _{1}+\alpha _{2})\cdot \tilde{q} &=&q_{13}\cosh \varepsilon +\frac{i%
}{\sqrt{3}}(q_{12}+q_{32})\sinh \varepsilon ,
\end{eqnarray}%
with the abbreviation $q_{ij}:=q_{i}-q_{j}$. We observe for this example the
following antilinear involutory symmetries 
\begin{eqnarray}
\mathcal{S}_{1} &:&\quad \quad q_{1}\leftrightarrow q_{2}\text{, }%
q_{3}\leftrightarrow q_{3}\text{, }i\rightarrow -i,  \label{S1} \\
\mathcal{S}_{2} &:&\quad \quad q_{2}\leftrightarrow q_{3}\text{, }%
q_{1}\leftrightarrow q_{1}\text{, }i\rightarrow -i.  \label{S2}
\end{eqnarray}

At this stage this deformation appears to be somewhat ad hoc. In fact, it
arose \cite{Milos,AFZ} from the physical motivation to eliminate
singularities in the potential when solving the separable Schr\"{o}dinger
equation for the Hamiltonian $\mathcal{H}_{\mathcal{PT}CMS}^{\prime }$. It
was noted that the new non-Hermitian model could be defined on less
separated configuration space. Whereas in general one had to restrict the
models to distinct Weyl chambers and analytically continue the wavefunctions
across their boundaries with the inclusion of some chosen phase, this is no
longer necessary in the deformed models. In addition, the new models possess
a modified energy spectrum with real eigenvalues, which we attribute to the
fact that the theory is invariant with respect to the antilinear
transformations $\mathcal{S}_{1}$ and $\mathcal{S}_{2}$. Motivated by this
success one may attempt to find a more direct systematic mathematical
procedure to deform the coordinates rather than the indirect implication
resulting from the separability of the Schr\"{o}dinger equation. In any
case, the latter approach would be entirely unpractical for models related
to higher rank Lie algebras.

We notice first that the Hamiltonian (\ref{defCMS}) also results from
deforming the roots involved. For the $A_{2}$-case we may take the simple
roots%
\begin{eqnarray}
\tilde{\alpha}_{1} &=&\alpha _{1}\cosh \varepsilon +i\frac{1}{\sqrt{3}}\sinh
\varepsilon (\alpha _{1}+2\alpha _{2}),  \label{al1} \\
\tilde{\alpha}_{2} &=&\alpha _{2}\cosh \varepsilon -i\frac{1}{\sqrt{3}}\sinh
\varepsilon (2\alpha _{1}+\alpha _{2}).  \label{al2}
\end{eqnarray}%
and re-write (\ref{defCMS}) equivalently as 
\begin{equation}
\mathcal{H}_{\mathcal{PT}CMS}^{\prime }=\frac{p^{2}}{2}+\frac{m^{2}}{16}%
\sum\limits_{\tilde{\alpha}\in \tilde{\Delta}_{s}}(\tilde{\alpha}\cdot
q)^{2}+\frac{1}{2}\sum\limits_{\tilde{\alpha}\in \tilde{\Delta}}g_{\tilde{%
\alpha}}V(\tilde{\alpha}\cdot q),\;~~~m,g_{\tilde{\alpha}}\in \mathbb{R}.
\label{defr}
\end{equation}%
Now the symmetries (\ref{S1})-(\ref{S2}) can be identified equivalently for
the roots. We note 
\begin{eqnarray}
\sigma _{1}^{\varepsilon } &:&\tilde{\alpha}_{1}\leftrightarrow -\tilde{%
\alpha}_{1}\text{, }\tilde{\alpha}_{2}\leftrightarrow \tilde{\alpha}_{1}+%
\tilde{\alpha}_{2}\quad \Leftrightarrow \quad q_{1}\leftrightarrow q_{2}%
\text{, }q_{3}\leftrightarrow q_{3}\text{, }i\rightarrow -i, \\
\sigma _{2}^{\varepsilon } &:&\tilde{\alpha}_{2}\leftrightarrow -\tilde{%
\alpha}_{2}\text{, }\tilde{\alpha}_{1}\leftrightarrow \tilde{\alpha}_{1}+%
\tilde{\alpha}_{2}\quad \Leftrightarrow \quad q_{2}\leftrightarrow q_{3}%
\text{, }q_{1}\leftrightarrow q_{1}\text{, }i\rightarrow -i.
\end{eqnarray}%
This observation has been taken as the basis for the formulation of a
systematic construction procedure leading to antilinerly invariant, and
therefore potentially physical, models \cite{Mon1,Mon2,Mon3}. The dynamical
variables, or possibly more general fields, appear in the dual space of some
roots with respect to the standard inner product. Since these root spaces
are naturally equipped with various symmetries due to the fact that by
construction they remain invariant under the action of the entire Weyl group 
$\mathcal{W}$, it is by far easier and systematic to identify the antilinear
symmetries directly in the root spaces rather than in the configuration
space. Once they have been identified they can be transformed to the latter.

The aim is therefore to construct complex extended antilinearly invariant
root systems which we denote by $\tilde{\Delta}(\varepsilon )$. The proposed
procedure consists of constructing two maps, which may be obtained in any
order. In one step we extend the representation space $\Delta $ of the
standard roots $\alpha $ from $\mathbb{R}^{n}$ to $\mathbb{C}^{n}=\mathbb{R}%
^{n}\oplus i\mathbb{R}^{n}$. This means we are seeking a map 
\begin{equation}
\delta :~\Delta \rightarrow \tilde{\Delta}(\varepsilon ),\qquad \alpha
\mapsto \tilde{\alpha}=\theta _{\varepsilon }\alpha ,  \label{defo}
\end{equation}%
where $\alpha =\{\alpha _{1},\ldots ,\alpha _{\ell }\}$, $\ \Delta \subset 
\mathbb{R}^{n}$, $\tilde{\Delta}(\varepsilon )\subset $ $\mathbb{C}^{n}$ and 
$n$ is greater or equal to the rank $\ell $ of the Weyl group $\mathcal{W}$.
The complex deformation matrix $\theta _{\varepsilon }$ introduced in (\ref%
{defo}) depends on the deformation parameter $\varepsilon $ in such a way
that $\lim_{\varepsilon \rightarrow 0}\theta _{\varepsilon }=\mathbb{I}$.
The deformation is constructed to facilitate the root space $\tilde{\Delta}$
with the crucial property for our purposes, namely to guarantee that it is
left invariant under an antilinear involutory map%
\begin{equation}
\varpi :\tilde{\Delta}(\varepsilon )\rightarrow \tilde{\Delta}(\varepsilon
),\qquad \tilde{\alpha}\mapsto \omega \tilde{\alpha}.  \label{map}
\end{equation}%
This means the map in (\ref{map}) satisfies $\varpi :\tilde{\alpha}=\mu
_{1}\alpha _{1}+\mu _{2}\alpha _{2}\mapsto \mu _{1}^{\ast }\omega \alpha
_{1}+\mu _{2}^{\ast }\omega \alpha _{2}$ for $\mu _{1}$, $\mu _{2}\in 
\mathbb{C}$ and also $\varpi \circ \varpi =\mathbb{I}$. In order to
facilitate the construction we make the further additional assumptions:

\begin{itemize}
\item[(i)] The operator $\omega $ decomposes as 
\begin{equation}
\omega =\tau \hat{\omega}=\hat{\omega}\tau ,
\end{equation}%
with $\hat{\omega}\in \mathcal{W}$, $\hat{\omega}^{2}=\mathbb{I}$ and $\tau $
being a complex conjugation. This will guarantee that $\varpi $ is
antilinear.

\item[(ii)] There are at least two different maps $\omega _{i}$ with $%
i=1,\ldots ,\kappa \geq 2$. This assumption simplifies the solution
procedure.

\item[(iii)] There exists a similarity transformation of the form 
\begin{equation}
\omega _{i}:=\theta _{\varepsilon }\hat{\omega}_{i}\theta _{\varepsilon
}^{-1}=\tau \hat{\omega}_{i},\qquad \text{for }i=1,\ldots ,\kappa \geq 2.
\end{equation}

\item[(iv)] The operator $\theta _{\varepsilon }$ is an isometry for the
inner products on $\tilde{\Delta}(\varepsilon )$, such that 
\begin{equation}
\theta _{\varepsilon }^{\ast }=\theta _{\varepsilon }^{-1}\qquad \text{ and}%
\qquad \det \theta _{\varepsilon }=\pm 1.
\end{equation}%
This assumption is motivated by the desire to keep the kinetic term of the
Calogero model undeformed.

\item[(v)] In the limit $\varepsilon \rightarrow 0$ we recover the
undeformed case%
\begin{equation}
\lim_{\varepsilon \rightarrow 0}\theta _{\varepsilon }=\mathbb{I}.
\end{equation}
\end{itemize}

Clearly one could modify or entirely relax some of the constraints (i)-(v),
e.g. it might not be desirable in some physical application to preserve the
inner products etc. However, it turns out that this set of constraints is
restrictive enough to allow for the construction of solutions for $\theta
_{\varepsilon }$ with only very few free parameters left.

With our applications to physical models in mind, i.e. exploiting here the
equivalence of (\ref{defCMS}) and (\ref{defr}), we would also like to
construct a dual map $\delta ^{\star }$ for $\delta $ acting on the
coordinate space with $q=\{q_{1},\ldots ,q_{n}\}$ or possibly fields. We
therefore define 
\begin{equation}
\delta ^{\star }:~\mathbb{R}^{n}\rightarrow \tilde{\Delta}^{\star
}(\varepsilon )=\mathbb{R}^{n}\oplus i\mathbb{R}^{n},\qquad x\mapsto \tilde{x%
}=\theta _{\varepsilon }^{\star }x,
\end{equation}%
denoting quantities in and acting on the dual space by $\star $. Thus
assuming $\theta _{\varepsilon }$ has been constructed from the constraints
(i)-(v), we may obtain $\theta _{\varepsilon }^{\star }$ by solving the $%
\ell $ equations 
\begin{equation}
(\tilde{\alpha}_{i}\cdot x)=(\left( \theta _{\varepsilon }\alpha \right)
_{i}\cdot x)=(\alpha _{i}\cdot \theta _{\varepsilon }^{\star }x)=(\alpha
_{i}\cdot \tilde{x}),\quad \text{for }i=1,\ldots ,\ell ,  \label{dual}
\end{equation}%
involving the standard inner product. This means $(\theta _{\varepsilon
}^{\star })^{-1}\alpha _{i}=\left( \theta _{\varepsilon }\alpha \right) _{i}$%
. Note that in general $\theta _{\varepsilon }^{\star }\neq \theta
_{\varepsilon }^{\ast }$. Naturally we can also identify an antilinear
involutory map 
\begin{equation}
\varpi ^{\star }:\tilde{\Delta}^{\star }(\varepsilon )\rightarrow \tilde{%
\Delta}^{\star }(\varepsilon ),\qquad \tilde{x}\mapsto \omega ^{\star }%
\tilde{x},  \label{dom}
\end{equation}%
corresponding to $\varpi $ but acting in the dual space. Concretely we need
to solve for this the $\kappa \times \ell $ relations%
\begin{equation}
\left( \omega _{i}\tilde{\alpha}\right) _{j}\cdot x=\alpha _{j}\cdot \omega
_{i}^{\star }\tilde{x},\quad \text{for }i=1,\ldots \kappa \text{; }%
j=1,\ldots ,\ell ,  \label{xc}
\end{equation}%
for $\omega _{i}^{\star }$ with given $\omega _{i}$.

In \cite{Mon1,Mon2,Mon3} many solutions to the set of constraints (i)-(v)
were constructed. A particular systematic construction can be found when we
take $\kappa =2$ in the requirement (ii) and identify $\omega _{1}=\sigma
_{-}$ and $\omega _{2}=\sigma _{+}$. The maps $\sigma _{\pm }$ factorize the
Coxeter element in a unique way 
\begin{equation}
\sigma :=\sigma _{-}\sigma _{+},\qquad \text{with }\sigma _{\pm
}:=\prod\limits_{i\in V_{\pm }}\sigma _{i},  \label{spm}
\end{equation}%
where the $\sigma _{i}$ are simple Weyl reflections $\sigma
_{i}(x):=x-2(x\cdot \alpha _{i}/\alpha _{i}^{2})\alpha _{i}$ associated to
each simple root for $1\leq i\leq \ell \equiv rank\mathcal{W}$. The two sets 
$V_{\pm }$ are defined by means of a bi-colouration of the Dynkin diagrams
consisting of associating values $c_{i}=\pm 1$ to its vertices in such a way
that no two vertices with the same values are linked together. The
consequence of this labeling is that $[\sigma _{i},\sigma _{j}]=0$ for $%
i,j\in V_{+}$ or $i,j\in V_{-}$ such that the factorization in (\ref{spm})
becomes unique. Clearly $\sigma _{\pm }^{2}=\mathbb{I}$ as required by for
our construction. An immediate consequence of (iii) is that $\sigma $ and $%
\theta _{\varepsilon }$ commute, such that the following Ansatz captures all
possible cases based on the assumption stated before (\ref{spm}) 
\begin{equation}
\theta _{\varepsilon }=\sum\limits_{k=0}^{h-1}c_{k}(\varepsilon )\sigma
^{k},\qquad \text{with }\lim_{\varepsilon \rightarrow 0}c_{k}(\varepsilon
)=\left\{ 
\begin{array}{l}
1\quad k=0 \\ 
0\quad k\neq 0%
\end{array}%
\right. ,~c_{k}(\varepsilon )\in \mathbb{C}.  \label{Ansatz}
\end{equation}%
The upper limit in the sum results from the fact that $\sigma ^{h}=\mathbb{I}
$, with $h$ denoting the Coxeter number. Invoking also the remaining
constraints allows to determine the functions $c_{k}(\varepsilon )$. For the 
$A_{3}$-Weyl group invariant root system this yields for instance the
following three deformed simple roots%
\begin{eqnarray}
\tilde{\alpha}_{1}\!\!\!\!\! &=&\!\!\!\!\!\cosh \varepsilon \alpha
_{1}+(\cosh \varepsilon -1)\alpha _{3}\!-\!i\sqrt{2}\sqrt{\cosh \varepsilon }%
\sinh \left( \frac{\varepsilon }{2}\right) \left( \alpha _{1}\!+\!2\alpha
_{2}\!+\!\alpha _{3}\right) , \\
\tilde{\alpha}_{2}\!\!\!\!\! &=&\!\!\!\!\!(2\cosh \varepsilon -1)\alpha
_{2}+2i\sqrt{2}\sqrt{\cosh \varepsilon }\sinh \left( \frac{\varepsilon }{2}%
\right) \left( \alpha _{1}+\alpha _{2}+\alpha _{3}\right) , \\
\tilde{\alpha}_{3}\!\!\!\!\! &=&\!\!\!\!\!\cosh \varepsilon \alpha
_{3}+(\cosh \varepsilon -1)\alpha _{1}\!-\!i\sqrt{2}\sqrt{\cosh \varepsilon }%
\sinh \left( \frac{\varepsilon }{2}\right) \left( \alpha _{1}\!+\!2\alpha
_{2}\!+\!\alpha _{3}\right) .~~
\end{eqnarray}%
In some cases we were even able to provide closed formulae for entire
subseries. For instance for $A_{4n-1}$ we found a closed expression for the
deformation matrix in the form 
\begin{equation}
\theta _{\varepsilon }=r_{0}\mathbb{I}+(1-r_{0})\sigma ^{2n}+i\sqrt{%
r_{0}^{2}-r_{0}}\left( \sigma ^{n}-\sigma ^{-n}\right) .
\end{equation}%
A possible choice for the function $r_{0}$ is $r_{0}=\cosh \varepsilon $. It
was also shown in \cite{Mon1} that it is impossible to construct solutions
for (i)-(v) for certain Weyl groups based on the factorization (\ref{spm}),
such as for instance $B_{2n+1}$. However, in \cite{Mon2,Mon3} it was
demonstrated that one can slightly alter the procedure by chosing different
factors instead and constructing solutions based on an Ansatz similar to (%
\ref{Ansatz}). For $B_{2n+1}$ we found closed expressions of the form 
\begin{eqnarray}
\tilde{\alpha}_{2j-1} &=&\cosh \varepsilon \alpha _{2j-1}+i\sinh \varepsilon
\left( \alpha _{2j-1}+2\sum\limits_{k=2j}^{\ell }\alpha _{k}\right) ~~~~~%
\text{for }j=1,\ldots ,n,  \label{a1} \\
\tilde{\alpha}_{2j} &=&\cosh \varepsilon \alpha _{2j}-i\sinh \varepsilon
\left( \sum\limits_{k=2j}^{2j+2}\alpha _{k}+2\sum\limits_{k=2j+3}^{\ell
}2\alpha _{k}\right) ~~\text{for }j=1,\ldots ,n-1,  \notag \\
\tilde{\alpha}_{\ell -1} &=&\cosh \varepsilon (\alpha _{\ell -1}+\alpha
_{\ell })-\alpha _{\ell }-i\sinh \varepsilon \left( \alpha _{\ell -2}+\alpha
_{\ell -1}+\alpha _{\ell }\right) ,  \notag \\
\tilde{\alpha}_{\ell } &=&\alpha _{\ell }.  \label{a4}
\end{eqnarray}%
In this case the dual deformation matrix which acts on the coordinates (\ref%
{dual}) takes on a very familiar form and turns out to be composed of
pairwise complex rotations%
\begin{equation}
\theta _{\varepsilon }^{\star }=\left( 
\begin{array}{ccccc}
R &  &  &  &  \\ 
& R &  & 0 &  \\ 
&  & \ddots &  &  \\ 
& 0 &  & R &  \\ 
&  &  &  & 1%
\end{array}%
\right) \qquad \text{with }R=\left( 
\begin{array}{rr}
\cosh \varepsilon & i\sinh \varepsilon \\ 
-i\sinh \varepsilon & \cosh \varepsilon%
\end{array}%
\right) .
\end{equation}

Having constructed various deformed root spaces which by construction are
equipped with an antilinear involutory symmetry, we may then consider
various models formulated in terms of roots, such as $\mathcal{H}_{\mathcal{%
PT}CMS}^{\prime }$ defined in (\ref{defr}). We then encounter several
interesting new features in these models. Since many of the key identities
needed for the solution procedure are identical in terms of roots or
deformed roots, we may adopt similar solution techniques as in the
undeformed case, such separating variables. As a crucial new feature we find
that the energy spectrum is modified and admits new real solutions when
compared to the undeformed model. For instance, for the $G_{2}$-case the
undeformed energies $E_{n}=2\left\vert \omega \right\vert (2n+\lambda +1)$
with $n\in \mathbb{N}_{0}$ and $\lambda \in \mathbb{R}^{+}$ become

\begin{equation}
E_{nm}^{\pm }=2|\omega |\left[ 2n+6(\kappa _{s}^{\pm }+\kappa _{l}^{\pm
}+m)+1\right] \qquad \text{for }n,m\in \mathbb{N}_{0},
\end{equation}%
with $\kappa _{s/l}^{\pm }=(1\pm \sqrt{1+4g_{s/l}})/4$.

For the $B_{2n+1}$-case we can support these observations with the explicit
construction of Dyson maps as introduced in (\ref{eta}) and the metric
operators (\ref{rel}). For the models based on the deformed roots (\ref{a1}%
)-(\ref{a4}) the Dyson map is simply $\eta =\eta _{12}\eta _{34}\ldots \eta
_{(\ell -2)(\ell -1)}$ with $\eta _{ij}=e^{-\varepsilon
(x_{i}p_{j}-x_{j}p_{i})}$, such that the metric operator becomes $\rho =\eta
^{2}$.

A further novelty in the deformed models is that the wavefunctions are
regularized by means of the deformation such that many singularities
disappear. In particular this means that these models can be defined usually
on the entire space $\mathbb{C}^{n}$, whereas the undeformed models could
only be defined in certain Weyl chambers. The continuation from a chamber to
its neighbouring one was achieved by introducing a phase factor by hand,
thus selecting a particular statistics. The deformed models on the other
hand have these phase factors already built in as a property of the model.
For instance in the $A_{3}$-case we find that the four-particle wavefunction
obeys 
\begin{equation}
\psi (q_{1},q_{2},q_{3},q_{4})=e^{\imath \pi s}\psi
(q_{2},q_{4},q_{1},q_{3}).  \label{anyon}
\end{equation}%
These properties can be read off easily from the action of the generalized $%
\mathcal{PT}$-symmetry on the deformed roots, translated to the dual space
that is to the coordinates and then to the parameterization of the
wavefunction. We note that the phase factor emerges as an intrinsic property
rather than as an imposition. We illustrate the relation (\ref{anyon}) as
follows

\unitlength=0.6000000pt 
\begin{picture}(300.0,70.00)(95.00,125.00)
\put(200.00,150.00){\circle*{10.00}}
\put(250.00,150.00){\circle*{10.00}}
\put(300.00,150.00){\circle*{10.00}}
\put(350.00,150.00){\circle*{10.00}}
\put(195.00,165.00){$ {\small w } $}
\put(245.00,165.00){$ {\small x } $}
\put(295.00,165.00){$ {\small y } $}
\put(345.00,165.00){$ {\small z } $}
\put(195.00,132.00){$ {\small q_{1}} $}
\put(245.00,132.00){$ {\small q_{2}} $}
\put(295.00,132.00){$ {\small q_{3}} $}
\put(345.00,132.00){$ {\small q_{4}} $}
\put(420.00,147.00){$ = \,\, e^{i \pi s} $}
\put(530.00,150.00){\circle*{10.00}}
\put(580.00,150.00){\circle*{10.00}}
\put(630.00,150.00){\circle*{10.00}}
\put(680.00,150.00){\circle*{10.00}}
\put(525.00,165.00){$ {\small w } $}
\put(575.00,165.00){$ {\small x } $}
\put(625.00,165.00){$ {\small y } $}
\put(675.00,165.00){$ {\small z } $}
\put(525.00,132.00){$ {\small q_{2}} $}
\put(575.00,132.00){$ {\small q_{4}} $}
\put(625.00,132.00){$ {\small q_{1}} $}
\put(675.00,132.00){$ {\small q_{3}} $}
\end{picture}

\noindent In the deformed models we can even allow some of the particles to
occupy the same place and scatter them with single particles

\unitlength=0.6000000pt 
\begin{picture}(300.0,70.00)(95.00,125.00)
\put(200.00,150.00){\circle*{10.00}}
\put(272.00,150.00){\circle*{10.00}}
\put(278.00,150.00){\circle*{10.00}}
\put(350.00,150.00){\circle*{10.00}}
\put(195.00,165.00){$ {\small x } $}
\put(270.00,165.00){$ {\small y } $}
\put(345.00,165.00){$ {\small z } $}
\put(250.00,132.00){$ {\small q_{2}=q_{3} } $}
\put(195.00,132.00){$ {\small q_{1}} $}
\put(345.00,132.00){$ {\small q_{4}} $}
\put(420.00,147.00){$ = \,\, e^{i \pi s} $}
\put(530.00,150.00){\circle*{10.00}}
\put(602.00,150.00){\circle*{10.00}}
\put(607.00,150.00){\circle*{10.00}}
\put(680.00,150.00){\circle*{10.00}}
\put(525.00,165.00){$ {\small x } $}
\put(600.00,165.00){$ {\small y } $}
\put(675.00,165.00){$ {\small z } $}
\put(523.00,132.00){$ {\small q_{2}  } $}
\put(580.00,132.00){$ {\small q_{1}=q_{4}  } $}
\put(673.00,132.00){$ {\small q_{3}  } $}
\end{picture}

\noindent We can also scatter pairs of particles resulting in the exchange
of one of the particles

\unitlength=0.6000000pt 
\begin{picture}(300.0,70.00)(95.00,125.00)
\put(197.00,150.00){\circle*{10.00}}
\put(203.00,150.00){\circle*{10.00}}
\put(347.00,150.00){\circle*{10.00}}
\put(353.00,150.00){\circle*{10.00}}
\put(195.00,165.00){$ {\small x } $}
\put(345.00,165.00){$ {\small y } $}
\put(173.00,132.00){$ {\small q_{1}=q_{2}} $}
\put(325.00,132.00){$ {\small q_{3}=q_{4}} $}
\put(420.00,147.00){$ = \,\, e^{i \pi s} $}
\put(527.00,150.00){\circle*{10.00}}
\put(533.00,150.00){ \circle*{10.00}}
\put(677.00,150.00){ \circle*{10.00}}
\put(683.00,150.00){ \circle*{10.00}}
\put(525.00,165.00){$ {\small x } $}
\put(675.00,165.00){$ {\small y } $}
\put(503.00,132.00){$ {\small q_{1}=q_{3}  } $}
\put(655.00,132.00){$ {\small q_{2}=q_{4}  } $}
\end{picture}

\noindent and we may even scatter triplets with a single particle

\unitlength=0.6000000pt 
\begin{picture}(300.0,70.00)(95.00,125.00)
\put(195.00,150.00){\circle*{10.00}}
\put(200.00,150.00){\circle*{10.00}}
\put(205.00,150.00){\circle*{10.00}}
\put(350.00,150.00){\circle*{10.00}}
\put(195.00,165.00){$ {\small x } $}
\put(345.00,165.00){$ {\small y } $}
\put(173.00,132.00){$ {\small q_{1}=q_{2}=q_{3} } $}
\put(345.00,132.00){$ {\small q_{4}} $}
\put(420.00,147.00){$ = $}
\put(530.00,150.00){\circle*{10.00}}
\put(675.00,150.00){\circle*{10.00}}
\put(680.00,150.00){\circle*{10.00}}
\put(685.00,150.00){\circle*{10.00}}
\put(525.00,165.00){$ {\small x } $}
\put(675.00,165.00){$ {\small y } $}
\put(523.00,132.00){$ {\small q_{4}  } $}
\put(635.00,132.00){$ {\small q_{1}=q_{2}=q_{3}} $}
\end{picture}

It is clear that these models have very interesting new properties and there
are still various open issues left worthwhile to investigate. More explicit
solutions for spectra and wavefunctions should be constructed; the important
questions of whether the deformed models are still integrable should be
settled; Dyson maps, the metric operators and Hermitian counterparts should
be constructed such that more observables of the models can be studied. The
constructed root systems could be used to formulate entirely new models of
different kind than Calogero systems.

\section{$\mathcal{PT}$-symmetrically deformed nonlinear wave equations}

The prototype integrable system of nonlinear wave type is the
Korteweg-deVries equation \cite{KdV} 
\begin{equation}
u_{t}+\beta uu_{x}+\gamma u_{xxx}=0,\quad \ \ \ \ \ \beta ,\gamma \in 
\mathbb{C},\ \ \ \ \   \label{KdV}
\end{equation}%
resulting from a Hamiltonian density%
\begin{equation}
\mathcal{H}_{\text{KdV}}[u]=-\frac{\beta }{6}u^{3}+\frac{\gamma }{2}%
u_{x}^{2}.\quad  \label{Hkdv}
\end{equation}%
The system admits two different types of $\mathcal{PT}$-symmetries%
\begin{equation}
\mathcal{PT}_{\pm }:x\mapsto -x,t\mapsto -t,i\mapsto -i,u\mapsto \pm u\quad
\ \ \text{for }\beta ,\gamma \in \mathbb{R},  \label{2PT}
\end{equation}%
which have only been exploited recently in \cite{BBCF,AFKdV,BBAF,CFB}.
According to the deformation prescription (\ref{defPT}) we can now deform
the Hamiltonian density (\ref{Hkdv}) in two alternative ways 
\begin{equation}
\delta _{\varepsilon }^{+}:u_{x}\mapsto u_{x,\varepsilon
}:=-i(iu_{x})^{\varepsilon }\qquad \text{or\qquad }\delta _{\varepsilon
}^{-}:u\mapsto u_{\varepsilon }:=-i(iu)^{\varepsilon },~~~~\ \ ~~
\label{delta}
\end{equation}%
respectively, depending on whether we assume $u(x,t)$ to be $\mathcal{PT}$%
-symmetric or $\mathcal{PT}$-anti-symmetric. The deformed models, suitably
normalised, are then defined by the densities%
\begin{equation}
\mathcal{H}_{\varepsilon }^{+}=-\frac{\beta }{6}u^{3}-\frac{\gamma }{%
1+\varepsilon }(iu_{x})^{\varepsilon +1},~~\text{and~~}\mathcal{H}%
_{\varepsilon }^{-}=\frac{\beta }{(1+\varepsilon )(2+\varepsilon )}%
(iu)^{\varepsilon +2}+\frac{\gamma }{2}u_{x}^{2},\quad  \label{Hpm}
\end{equation}%
with corresponding equations of motions%
\begin{equation}
u_{t}+\beta uu_{x}+\gamma (u_{x,\varepsilon })_{xx}=0,\qquad \text{and\qquad 
}u_{t}+i\beta u_{\varepsilon }u_{x}+\gamma u_{xxx}=0.  \label{dIto2}
\end{equation}%
The $\mathcal{PT}$-symmetry can be exploited to ensure the reality for
expressions such as the energy on a certain interval $[-a,a]$ 
\begin{equation}
E=\int\nolimits_{-a}^{a}\mathcal{H}\left[ u(x)\right] dx=\oint\nolimits_{%
\Gamma }\mathcal{H}\left[ u(x)\right] \frac{du}{u_{x}}.  \label{Energy}
\end{equation}%
One would expect this expression to be real for the unbroken symmetric
regime. However, in \cite{CFB} also some unexpected cases with real energies
were found for which the $\mathcal{PT}$-symmetry is entirely broken (\ref%
{2PT}), i.e. for the Hamiltonian and for its solutions. This possibly
indicates the existence of a different realisation for the $\mathcal{PT}$-symmetry operator.

\begin{figure}[h]
\centering\includegraphics[width=6.5cm]{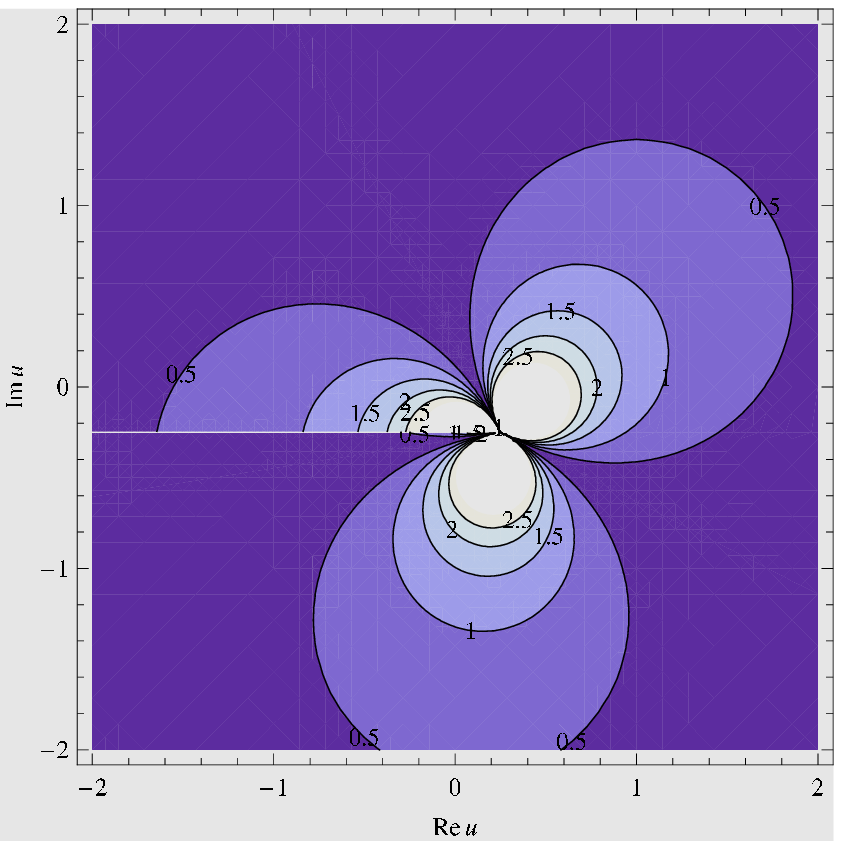} %
\includegraphics[width=6.5cm]{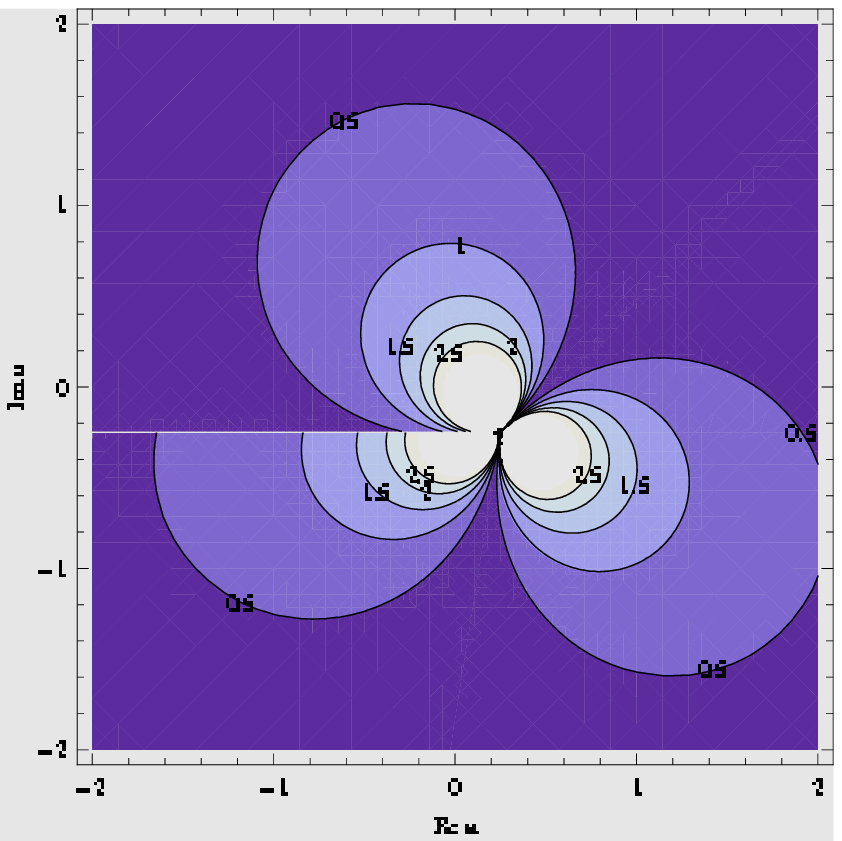}
\caption{The two different Riemann sheets for rational solutions with broken 
$\mathcal{PT}$-symmetry of the $\mathcal{H}_{1/3}^{+}$-model with different
values of purely complex initial conditions whose imaginary part is
indicated in the figure $\ $The asymptotic conditions are chosen to be $%
\lim_{\protect\zeta \rightarrow \pm \infty }u(\protect\zeta )=(1-i)/4$, the
speed of the wave as $c=1$ and the constants of the model as $\protect\beta %
=2+2i$ and $\protect\gamma =3$ }
\label{fig2}
\end{figure}

A further characteristic feature of the deformed models is that, in general,
one has to view them on various Riemann sheets. An example for a traveling
wave solution parameterized by $\zeta =x-ct$, with $c$ denoting the wave
speed, for broken $\mathcal{PT}$-symmetry is presented in figure \ref{fig2}.

The branch cut at $-\infty -i/4$ to $(1-i)/4$ is passed from above in panel
(a) to below in panel (b). The trajectories for the $\mathcal{PT}$-symmetric
and broken $\mathcal{PT}$-symmetric case look qualitatively very similar,
the major difference being that the fixed point has moved away from the real
axis, thus leading to a loss of the symmetry.

\begin{figure}[h]
\centering\includegraphics[width=4.2cm]{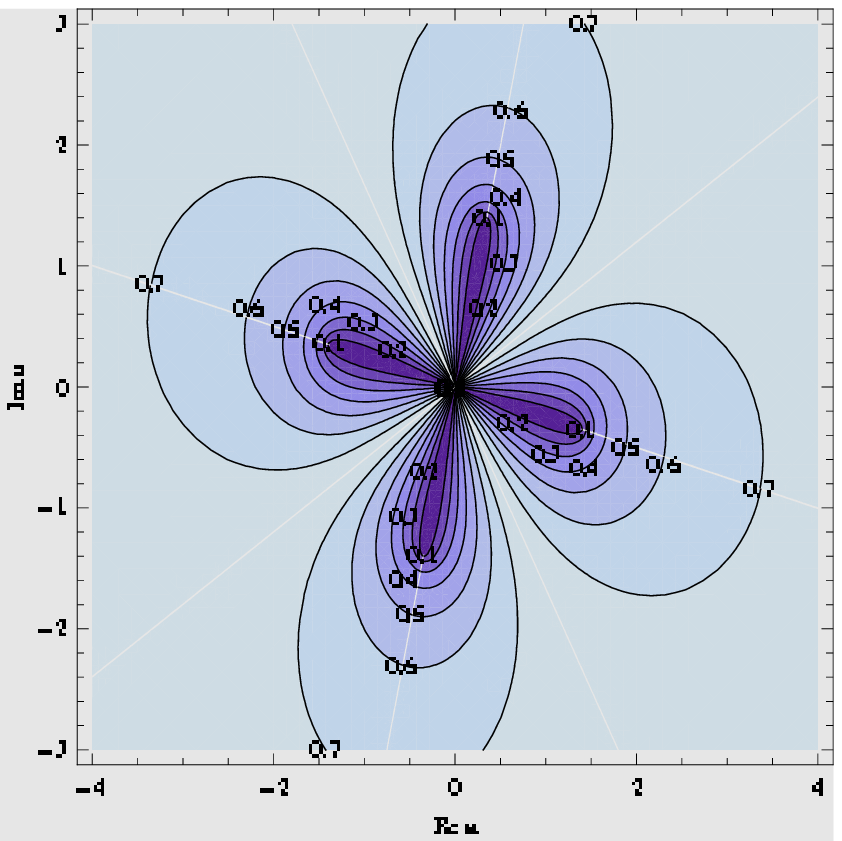} %
\includegraphics[width=4.2cm]{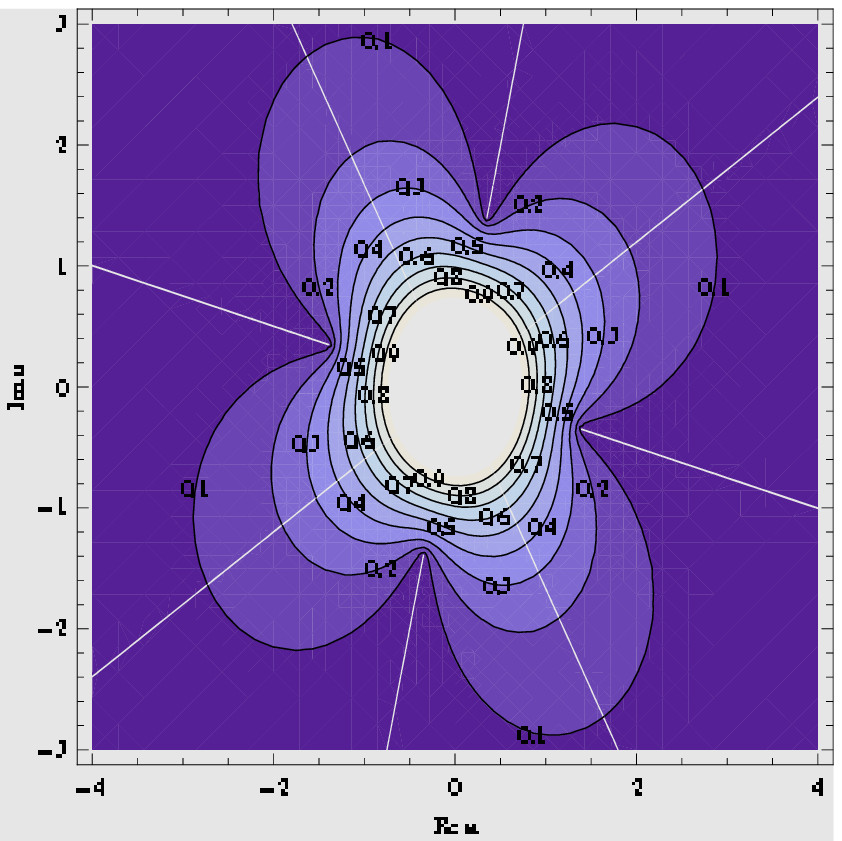}%
\includegraphics[width=4.2cm]{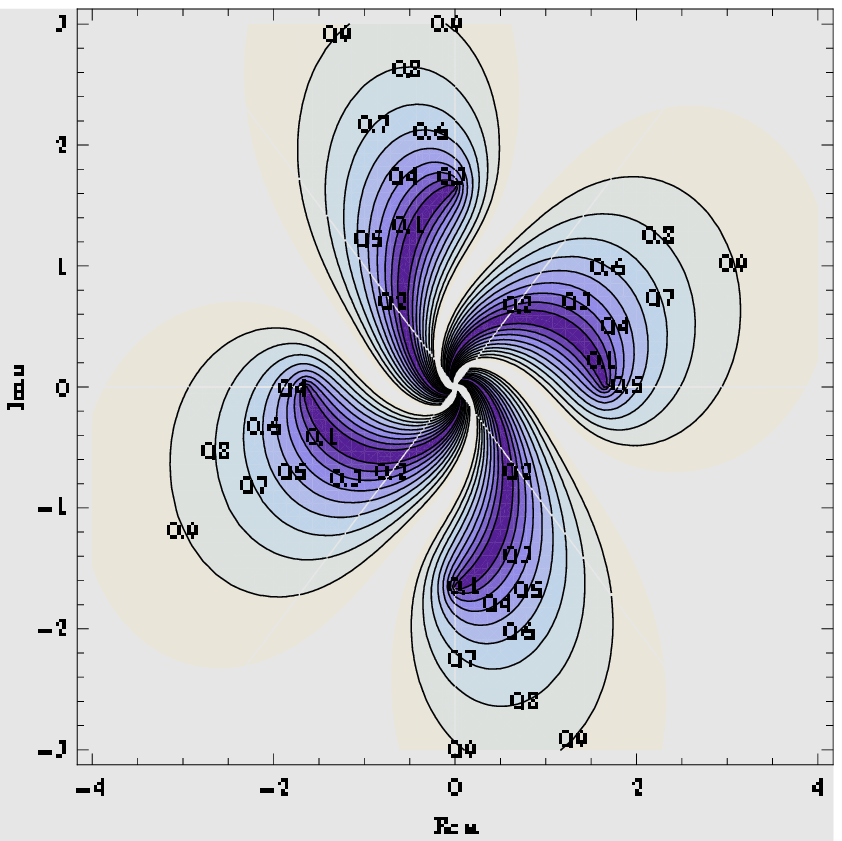}
\caption{Broken $\mathcal{PT}$-symmetric solution for $\mathcal{H}_{4}^{-}$:
(a) Star node at the origin for $c=1$, $\protect\beta =2+i3$, $\protect%
\gamma =1$; (b) centre at the origin for $c=1$, $\protect\beta =2+i3$, $%
\protect\gamma =-1$; (c) focus at the origin for $c=1$, $\protect\beta =2$, $%
\protect\gamma =1+i3$.}
\label{fig3}
\end{figure}

Viewing the systems as two dimensional models, the nature of the fixed
points has been investigated systematically by exploiting the fact that
their characteristic behaviour is completely classified in dependence on the
different types of eigenvalues for the Jacobian. In \cite{CFB} it was found
that they may even undergo Hopf bifurcations in these systems, passing form
a star node over a centre to a focus. This feature was derived for the $%
\mathcal{PT}$-symmetric as well as for the broken $\mathcal{PT}$-symmetric
regime. In particular this also means that we encounter closed trajectories
despite the fact that the $\mathcal{PT}$-symmetry is broken. We depict an
examples in figure \ref{fig3} for different values of $c$, $\beta $ and $%
\gamma $.

An interesting relation between these type of deformations and some simple
complex quantum mechanical models was pointed out in \cite{CFB}. As a
special case of this general observation we consider here the model $%
\mathcal{H}_{2}^{-}$-model with Hamiltonian density%
\begin{equation}
\mathcal{H}_{2}^{-}[u]=\frac{\beta }{12}u^{4}+\frac{\gamma }{2}u_{x}^{2},
\label{Hm2}
\end{equation}%
and make contact with the model studied in \cite{Anderson}. As explained in 
\cite{CFB}, integrating (\ref{dIto2}) twice with respect to $\zeta $ we
obtain 
\begin{equation}
u_{\zeta }^{2}=\frac{2}{\gamma }\left( \kappa _{2}+\kappa _{1}u+\frac{c}{2}%
u^{2}+\beta \frac{1}{12}u^{4}\right) ,  \label{po}
\end{equation}%
with integration constants $\kappa _{1},\kappa _{2}\in \mathbb{C}$.
Identifying $u\rightarrow x$ and $\zeta \rightarrow t$ for the traveling
wave equation together with the constraints 
\begin{equation}
\kappa _{1}=-\gamma \tau ,\quad \kappa _{2}=\gamma E_{x},\quad \beta
=-3\gamma g\quad \text{and\quad }c=-\gamma \omega ^{2},  \label{ch}
\end{equation}%
converts the derivative of equation (\ref{po}) into Newton's equation 
\begin{equation}
\ddot{x}+\tau +\omega ^{2}x+gx^{3}=0,  \label{xx}
\end{equation}%
for the quartic harmonic oscillator of the form%
\begin{equation}
H_{\text{quartic}}=E_{x}=\frac{1}{2}p^{2}+\tau x+\frac{\omega ^{2}}{2}x^{2}+%
\frac{g}{4}x^{4}.  \label{4}
\end{equation}%
One may now directly translate some of the properties of the system (\ref%
{Hm2}) to the quantum mechanical model (\ref{4}). The special choice $\kappa
_{1}=\tau =0$ for the integrations constants imply that one is considering
asymptotically vanishing waves with $\lim_{\zeta \rightarrow \infty }u(\zeta
)=0$ and with Neumann boundary condition $\lim_{\zeta \rightarrow \infty
}u_{x}(\zeta )=\sqrt{2E_{x}}$ where $H_{\text{quartic}}=E_{x}$. Accordingly,
the energy in the classical analogue of a complex classical particle
corresponds to an integration constant in the nonlinear wave equation
context multiplied by one of the coupling constants in the latter model.
This is of course different from the energy as defined in (\ref{Energy}),
which also leads to different conclusions regarding the reality of these
quantities resulting from the various $\mathcal{PT}$-symmetric scenarios.

In a similar way, the complex seminal \cite{Bender:1998ke} cubic harmonic
oscillator 
\begin{equation}
H_{\text{cubic}}=\frac{1}{2}p^{2}+\frac{1}{2}x^{2}+igx^{3},
\end{equation}%
treated also in \cite{Bender:2008fr} simply results from the integrating the
KdV-equation twice with the identification $\kappa _{1}=0$, $\kappa
_{2}=\gamma E_{x}$, $\beta =-i6cg$ and $\gamma =-c$.

It appears to be unlikely that the models are still integrable as in general
they do not pass the Painlev\'{e} test \cite{PainAF,CompactonsAF}. Similar
studies have also be carried out for other types of nonlinear wave equations
as for instance for deformed Ito systems in \cite{CFB}. It was even shown
that one can $\mathcal{PT}$-symmetrically deform the supersymmetric version
of the KdV-equation (\ref{KdV}) while still preserving its supersymmetry 
\cite{BBAF}.

Evidently many features remain still unexplored and it would be very
interesting to extend these studies to a larger range of values for the
deformation parameter, to other nonlinear field equations such as Burgers,
Bussinesque, KP, generalized shallow water equations, extended KdV equations
with compacton solution, etc.

\medskip

\noindent \textbf{Acknowledgments:} I would like to thank all my
collaborators on the topic presented here for sharing their insights: Paulo
Assis, Bijan Bagchi, Olalla Castro-Alvaredo, Andrea Cavaglia, Sanjib Dey,
Carla Figueira de Morisson Faria, Laure Gouba, Frederik Scholtz, Monique
Smith and Miloslav Znojil.


\begin{thebibliography}{10}

\bibitem{Bender:1998ke}
C.~M. Bender and S.~Boettcher,
\newblock Real Spectra in Non-Hermitian Hamiltonians Having PT Symmetry,
\newblock Phys. Rev. Lett. {\bf 80}, 5243--5246 (1998).

\bibitem{Urubu}
F.~G. Scholtz, H.~B. Geyer, and F.~Hahne,
\newblock Quasi-Hermitian Operators in Quantum Mechanics and the Variational
  Principle,
\newblock Ann. Phys. {\bf 213}, 74--101 (1992).

\bibitem{Bender:2002vv}
C.~M. Bender, D.~C. Brody, and H.~F. Jones,
\newblock Complex Extension of Quantum Mechanics,
\newblock Phys. Rev. Lett. {\bf 89}, 270401(4) (2002).

\bibitem{Mostafazadeh:2002hb}
A.~Mostafazadeh,
\newblock Pseudo-Hermiticity versus PT symmetry. The necessary condition for
  the reality of the spectrum,
\newblock J. Math. Phys. {\bf 43}, 205--214 (2002).

\bibitem{Bender:2003ve}
C.~M. Bender, D.~C. Brody, and H.~F. Jones,
\newblock Must a Hamiltonian be Hermitian?,
\newblock Am. J. Phys. {\bf 71}, 1095--1102 (2003).

\bibitem{Mostafazadeh:2004mx}
A.~Mostafazadeh and A.~Batal,
\newblock Physical Aspects of Pseudo-Hermitian and $PT$-Symmetric Quantum
  Mechanics,
\newblock J. Phys. {\bf A37}, 11645--11680 (2004).

\bibitem{Caliceti:2004xw}
E.~Caliceti, F.~Cannata, M.~Znojil, and A.~Ventura,
\newblock Construction of PT-asymmetric non-Hermitian Hamiltonians with
  CPT-symmetry,
\newblock Phys. Lett. {\bf A335}, 26--30 (2005).

\bibitem{CA}
C.~Figueira~de Morisson~Faria and A.~Fring,
\newblock Time evolution of non-Hermitian Hamiltonian systems,
\newblock J. Phys. {\bf A39}, 9269--9289 (2006).

\bibitem{Moyal1}
F.~G. Scholtz and H.~B. Geyer,
\newblock Operator equations and Moyal products -- metrics in quasi-hermitian
  quantum mechanics,
\newblock Phys. Lett. {\bf B634}, 84--92 (2006).

\bibitem{ACIso}
C.~Figueira~de Morisson~Faria and A.~Fring,
\newblock Isospectral Hamiltonians from Moyal products,
\newblock Czech. J. Phys. {\bf 56}, 899--908 (2006).

\bibitem{Mostsyme}
A.~Mostafazadeh,
\newblock Metric operators for quasi-Hermitian Hamiltonians and symmetries of
  equivalent Hermitian Hamiltonians,
\newblock J. Phys. {\bf A41}, 055304 (2008).

\bibitem{EW}
E.~Wigner,
\newblock Normal form of antiunitary operators,
\newblock J. Math. Phys. {\bf 1}, 409--413 (1960).

\bibitem{Dieu}
J.~Dieudonn{\'{e}},
\newblock Quasi-hermitian operators,
\newblock Proceedings of the International Symposium on Linear Spaces,
  Jerusalem 1960, Pergamon, Oxford , 115--122 (1961).

\bibitem{Will}
J.~P. Williams,
\newblock Operators similar to their adjoints,
\newblock Proc. American. Math. Soc. {\bf 20}, 121--123 (1969).

\bibitem{pseudo1}
M.~Froissart,
\newblock Covariant formalism of a field with indefinite metric,
\newblock Il Nuovo Cimento {\bf 14}, 197--204 (1959).

\bibitem{pseudo2}
E.~C.~G. Sudarshan,
\newblock Quantum Mechanical Systems with Indefinite Metric. I,
\newblock Phys. Rev. {\bf 123}, 2183--2193 (1961).

\bibitem{DDT}
P.~Dorey, C.~Dunning, and R.~Tateo,
\newblock Spectral equivalences from Bethe ansatz equations,
\newblock J. Phys. {\bf A34}, 5679--5704 (2001).

\bibitem{CFB}
A.~Cavaglia, A.~Fring, and B.~Bagchi,
\newblock PT-symmetry breaking in complex nonlinear wave equations and their
  deformations,
\newblock J. Phys. {\bf A44}, 325201 (2011).

\bibitem{Guenter1}
G.~von Gehlen,
\newblock Critical and off-critical conformal analysis of the Ising quantum
  chain in an imaginary field,
\newblock J. Phys. {\bf A24}, 5371--5399 (1991).

\bibitem{Cardy:1985yy}
J.~L. Cardy,
\newblock Conformal invariance and the Yang-Lee edge singularity in
  two-dimension,
\newblock Phys. Rev. Lett. {\bf 54}, 1354--1356 (1985).

\bibitem{BPZ}
A.~A. Belavin, A.~M. Polyakov, and A.~B. Zamolodchikov,
\newblock Infinite conformal symmetry in two-dimensional quantum field theory,
\newblock Nucl. Phys. {\bf B241}, 333--380 (1984).

\bibitem{chainOla}
O.~A. Castro-Alvaredo and A.~Fring,
\newblock A spin chain model with non-Hermitian interaction: The Ising quantum
  spin chain in an imaginary field,
\newblock J. Phys. {\bf A42}, 465211 (2009).

\bibitem{CKW}
C.~Korff and R.~A. Weston,
\newblock PT Symmetry on the Lattice: The Quantum Group Invariant XXZ
  Spin-Chain,
\newblock J. Phys. {\bf A40}, 8845--8872 (2007).

\bibitem{DeGh}
T.~D. and P.~K. G.,
\newblock The exactly solvable quasi-Hermitian transverse Ising model,
\newblock J. Phys. {\bf A42}, 475208 (2009).

\bibitem{CarlExPoint}
C.~M. Bender and T.~T. Wu,
\newblock Anharmonic Oscillator,
\newblock Phys. Rev. {\bf 184}, 1231--1260 (1969).

\bibitem{vNW}
J.~von Neuman and E.~Wigner,
\newblock {\"U}ber merkw{\"u}rdige diskrete Eigenwerte. {\"U}ber das Verhalten
  von Eigenwerten bei adiabatischen Prozessen,
\newblock Zeit. der Physik {\bf 30}, 467--470 (1929).

\bibitem{Dyson}
F.~J. Dyson,
\newblock Thermodynamic Behavior of an Ideal Ferromagnet,
\newblock Phys. Rev. {\bf 102}, 1230--1244 (1956).

\bibitem{Andrei}
A.~G. Bytsko,
\newblock Non-Hermitian spin chains with inhomogeneous coupling,
\newblock Phys. Rev. B {\bf 22N3}, 80--106 (2010).

\bibitem{Georgi}
G.~L. Giorgi,
\newblock Spontaneous $\mathcal{P}\mathcal{T}$ symmetry breaking and quantum
  phase transitions in dimerized spin chains,
\newblock Phys. Rev. B {\bf 82}, 052404 (2010).

\bibitem{AF}
A.~Fring,
\newblock A note on the integrability of non-Hermitian extensions of
  Calogero-Moser-Sutherland models,
\newblock Mod. Phys. Lett. {\bf 21}, 691--699 (2006).

\bibitem{Basu-Mallick:2001ce}
B.~Basu-Mallick and B.~P. Mandal,
\newblock On an exactly solvable $B_N$ type Calogero model with nonhermitian PT
  invariant interaction,
\newblock Phys. Lett. {\bf A284}, 231--237 (2001).

\bibitem{Basu-Mallick:2003pt}
B.~Basu-Mallick, T.~Bhattacharyya, A.~Kundu, and B.~P. Mandal,
\newblock Bound and scattering states of extended Calogero model with an
  additional PT invariant interaction,
\newblock Czech. J. Phys. {\bf 54}, 5--12 (2004).

\bibitem{Basu-Mallick:2004ye}
B.~Basu-Mallick, T.~Bhattacharyya, and B.~P. Mandal,
\newblock Phase shift analysis of PT-symmetric nonhermitian extension of
  $A_{N-1}$ Calogero model without confining interaction,
\newblock Mod. Phys. Lett. {\bf A20}, 543--552 (2005).

\bibitem{OP2}
M.~A. Olshanetsky and A.~M. Perelomov,
\newblock Classical integrable finite dimensional systems related to Lie
  algebras,
\newblock Phys. Rept. {\bf 71}, 313--400 (1981).

\bibitem{Chen}
H.~H. Chen, L.~Y. C., and N.~R. Pereira,
\newblock Algebraic internal wave solutions and the integrable
  Calogero-Moser-Sutherland $N$-body problem,
\newblock Phys. Fluids {\bf 22}, 187--188 (1979).

\bibitem{JoshLondon}
J.~Feinberg,
\newblock talk at PHHQP workshop VI, City University London,
\newblock (2007).

\bibitem{AMM}
H.~Airault, H.~P. McKean, and J.~Moser,
\newblock Rational and Elliptic Solutions of the Korteweg-de Vries Equation and
  a Related Many-Body Problem,
\newblock Comm. pure appl. Math. {\bf 80}, 95--148 (1977).

\bibitem{Assis:2009gt}
P.~E.~G. Assis and A.~Fring,
\newblock From real fields to complex Calogero particles,
\newblock J. Phys. {\bf A42}, 425206(14) (2009).

\bibitem{Milos}
M.~Znojil and M.~Tater,
\newblock Complex Calogero model with real energies,
\newblock J. Phys. {\bf A34}, 1793--1803 (2001).

\bibitem{AFZ}
A.~Fring and M.~Znojil,
\newblock $\cal{PT}$-Symmetric deformations of Calogero models,
\newblock J. Phys. {\bf A40}, 194010(17) (2008).

\bibitem{Mon1}
A.~Fring and M.~Smith,
\newblock Antilinear deformations of Coxeter groups, an application to Calogero
  models,
\newblock J. Phys. {\bf A43}, 325201 (2010).

\bibitem{Mon2}
A.~Fring and M.~Smith,
\newblock PT invariant complex E(8) root spaces,
\newblock Int. J. Theor. Phys. {\bf 50}, 974--981 (2011).

\bibitem{Mon3}
A.~Fring and M.~Smith,
\newblock Non-Hermitian multi-particle systems from complex root spaces,
\newblock J. Phys. {\bf A45}, 085203 (2012).

\bibitem{KdV}
D.~J. Korteweg and deVries G.,
\newblock On the change of form of long waves advancing in a rectangular canal,
  and on a new type of long stationary waves,
\newblock Phil. Mag. {\bf 39}, 422--443 (1895).

\bibitem{BBCF}
C.~M. Bender, D.~C. Brody, J.~Chen, and E.~Furlan,
\newblock $\cal{PT}$-symmetric extension of the Korteweg-de Vries equation,
\newblock J. Phys. {\bf A40}, F153--F160 (2007).

\bibitem{AFKdV}
A.~Fring,
\newblock $\cal{PT}$-Symmetric deformations of the Korteweg-de Vries equation,
\newblock J. Phys. {\bf A40}, 4215--4224 (2007).

\bibitem{BBAF}
B.~Bagchi and A.~Fring,
\newblock $\cal{PT}$-symmetric extensions of the supersymmetric Korteweg-De
  Vries equation,
\newblock J. Phys. {\bf A41}, 392004(9) (2008).

\bibitem{Anderson}
A.~G. Anderson, C.~M. Bender, and U.~I. Morone,
\newblock Periodic orbits for classical particles having complex energy,
\newblock Phys. Lett. {\bf A375}, 3399–3404 (2011).

\bibitem{Bender:2008fr}
C.~M. Bender, D.~C. Brody, and D.~W. Hook,
\newblock Quantum effects in classical systems having complex energy,
\newblock J. Phys. {\bf A41}, 352003 (2008).

\bibitem{PainAF}
P.~E.~G. Assis and A.~Fring,
\newblock Integrable models from -symmetric deformations,
\newblock J. Phys. {\bf A42}, 105206 (2009).

\bibitem{CompactonsAF}
P.~E.~G. Assis and A.~Fring,
\newblock Compactons versus Solitons,
\newblock Pramana J. Phys. {\bf 74}, 857--865 (2010).

\end{thebibliography}

\end{document}